\documentclass[twocolumn]{aastex61}
\usepackage{amsmath}
\usepackage[caption=false]{subfig}
\usepackage{graphicx}
\usepackage{epstopdf} 
\usepackage{float}
\usepackage[normalem]{ulem}
\usepackage{subfig}
\usepackage{gensymb}
\usepackage{units}

\submitjournal{ApJ}

\begin{document}

\title{Numerical simulations of the jet dynamics and synchrotron radiation
  of %% of the afterglow of 
  binary neutron star merger event GW170817/GRB170817A}

\correspondingauthor{Xiaoyi Xie}
\email{Email: xiaoyi@nyu.edu, macfadyen@nyu.edu}
%\email{macfadyen@nyu.edu}

\author{Xiaoyi Xie}
\affiliation{Center for Cosmology and Particle Physics, Physics Department New York University, 726 Broadway, New York, NY 10003, USA}
\author{Jonathan Zrake}
\affiliation{Columbia University, Pupin Hall, 550 West 120th Street, New
  York, NY 10027, USA}
\author[0000-0002-0106-9013]{Andrew MacFadyen}
\affiliation{Center for Cosmology and Particle Physics, Physics Department New York University, 726 Broadway, New York, NY 10003, USA}

% \collaboration{(AAS Journals Data Scientists collaboration)}



\begin{abstract}

  We present numerical simulations of energetic flows
  propagating through the debris cloud of a binary neutron star (BNS)
  merger. Starting from the scale of the central engine, we use a
  moving-mesh hydrodynamics code to simulate the complete dynamical evolution of the produced relativistic jets.  We compute synchrotron
  emission directly from the simulations and present multi-band light
  curves of the early (sub-day) through late (weeks to years)
  afterglow stages. Our work systematically compares two distinct
  models for the central engine, referred to as the narrow and wide
  engine scenario, which is associated with a successful structured
  jet and a quasi-isotropic explosion respectively. Both engine models
  naturally evolve angular and radial structure through hydrodynamical
  interaction with the merger debris cloud. They both also result in a
  relativistic blast wave capable of producing the observed multi-band
  afterglow data. However, we find that the narrow and wide engine scenario
  might be differentiated by a new emission component that we refer to
  as a \emph{merger flash}. This component is a consequence of
    applying the synchrotron radiation model to the shocked optically thin merger cloud. Such modeling is appropriate if injection of non-thermal electrons is sustained in the breakout relativistic shell, for example by internal shocks or magnetic reconnection. The rapidly declining signature may be detectable for future BNS mergers during the first minutes to day
  following the GW chirp. Furthermore, its non-detection for the
  GRB170817A event may disfavor the wide,
  quasi-isotropic explosion model.

% We show that in the narrow engine model, current afterglow observations already detect emission from the jet core, suggesting that the afterglow light curve of GRB170817A will not rise significantly 

% In the first scenario, a relativistic jet successfully breaks out of the merger cloud, is directed away from our line-of-sight by roughly $20^\circ$, and has significant angular structure. We show that the jet naturally acquires the requisite structure through its hydrodynamical interaction with the merger cloud. In the second scenario, a relatively isotropic outflow of mildly relativistic material emerges from the merger cloud as the result of a failed jet (one fails to continue propagating beyond breakout). Both scenarios predict that the non-thermal afterglow emission of GRB170817A decays at $\sim 200\,\,\rm{days}$ after the BNS merger, and will remain detectable by \emph{Chandra} for at least $30$ years. In future multi-messenger detections of BNS mergers, these two scenarios may be differentiated by the nature of their cocoon emission. We discuss the possibility that the cooling emission of the relativistic cocoon material is non-thermal, and show corresponding sub-day light-curves. An early declining X-ray (rather than UV) transient, associated with the relativistic cocoon, may be detectable by current or proposed X-ray observatories.

\end{abstract}

% \keywords{Short Gamma-Ray Burst, Relativistic Hydrodynamics, High Energy Astrophysics}



\section{Introduction} \label{sec:intro}
On August 17, 2017 the Laser Interferometer Gravitational Wave Observatory (LIGO) detected the first gravitational wave (GW) signal from the merger of a binary neutron star system \citep{2017ApJ...848L..12A}. About 1.7 s later, the Fermi Gamma-ray Burst Monitor (GBM) detected a coincident short Gamma-Ray Burst (sGRB), marking the first confident joint electromagnetic (EM)-gravitational wave (GW) observation in history \citep{2017ApJ...848L..14G, 2017ApJ...848L..15S}. Follow-up observing campaigns across the electromagnetic spectrum were launched to discover the merger site and observe its ongoing electromagnetic emission. In less than 11 hours after the merger, a bright optical transient was discovered in the galaxy NGC4993 (at $\sim \unit[40]{Mpc}$) \citep{2017Sci...358.1556C, 2017ApJ...848L..12A, 2017ApJ...848L..16S, 2017ApJ...848L..24V}. Early ultraviolet-optical-infrared (UVOIR) data from multiple telescopes throughout the world reveal a quasi-thermal radiation component, which is consistent with the prediction of the ``kilonova/macronova'' model \citep{2017LRR....20....3M, 2017ApJ...848L..19C, 2017ApJ...848L..17C,2017ApJ...848L..18N, 2017Natur.551...75S,2017ApJ...848L..16S, 2017ApJ...848L..27T, 2017ApJ...851L..21V, 2017Sci...358.1559K, 2017Natur.551...80K, 2017Natur.551...67P,2017arXiv171109638W}. However, no X-ray and radio signals were detected during the first several days, though important flux upper-limits were obtained \citep{2017ApJ...848L..12A}. The first detection of X-rays came from \emph{Chandra} 9 days after the GW event \citep{2017ApJ...848L..20M, 2017Natur.551...71T}. Radio emission was first detected 16 days after the GW event \citep{2017ApJ...848L..21A}.

Continuous X-ray and radio observations show steadily increasing luminosity up to
$\sim \unit[100] {days}$ after the GW event \citep{2017ApJ...848L..25H, 2017Sci...358.1579H, 2018Natur.554..207M, 2018ApJ...853L...4R}. Recent observations ($\sim \unit[200]{days}$ after GW) may show a hint of a turn-over in the radio, optical and X-ray light curves \citep{2018arXiv180102669L, 2018ApJ...856L..18M, 2018arXiv180106164D, 2018arXiv180306853D}.

The late-time non-thermal light curves can be interpreted as
synchrotron emission from a relativistic blast wave that is launched from the
merger and propagating in the circum-merger environment (hereafter,
referred to as the interstellar medium (ISM)). Several scenarios have
been proposed for describing the blast wave launched from the merging process.
On-axis and off-axis ``top-hat'' Blandford-McKee (BM)
\citep{1976PhFl...19.1130B} jet models, for which the energy and
radial velocity are uniform within a cone and drop discontinuously to
zero outside of the cone, have been ruled out (see
\citealt{2018Natur.554..207M,2017Sci...358.1559K}). Two dynamical
models for the central engine remain under consideration: 1) a narrow
ultra-relativistic engine that produces a successful jet with
  angular structure and 2) a wide trans-relativistic engine that
produces a quasi-spherical mildly
  relativistic outflow. The first scenario has been referred to as a
``successful structured jet'' %% e.g., \citealt{2018ApJ...856L..18M}
(
\citealt{2018MNRAS.tmp.1056L,2017arXiv171203237L,2017Natur.551...71T,2018arXiv180106164D,2018ApJ...856L..18M,2018arXiv180302768R,2018MNRAS.tmpL..60T,2018MNRAS.tmp.1159G}),
  and the second has been referred to variously as a ``choked jet'' or
  a ``failed jet'' 
    \citep{2017Sci...358.1559K,2018arXiv180106164D,2017arXiv171005896G,2018MNRAS.473..576G,2018Natur.554..207M,
      2018arXiv180307595N,2018MNRAS.tmpL..60T}. The
    non-thermal emission of the proposed jet profiles has been
    calculated analytically
    (e.g. \citealt{2018MNRAS.tmp.1056L,2018arXiv180106164D,2018Natur.554..207M,2018arXiv180106516T,2018MNRAS.tmp.1159G,2017ApJ...848L..34M})
    or semi-analytically with numerical simulations covering the
    initial phases of the jet evolution (e.g
    \citealt{2017arXiv171203237L,2017Sci...358.1559K,2017arXiv171005896G}). \cite{2018arXiv180307595N}
    has carried out simulations utilizing the PLUTO code
    \citep{2007ApJS..170..228M} and calculated the full-EM emission from the mildly relativistic cocoon.

In this work, we utilize the moving-mesh relativistic hydrodynamics
code JET \citep{2013ApJ...775...87D} to conduct
high-resolution full-time-domain simulations of these two dynamical
models, starting at the scale of the central engine and evolving
continuously to the scale of the afterglow. Broad-band light curves
computed from these simulations, utilizing a well-tested
synchrotron radiation code \citep{2009ApJ...698.1261Z,2010ApJ...722..235V}, can be used to interpret data from
future BNS mergers, which are expected to occur at rates of up to 1/month after advanced LIGO and Virgo commence operation in early 2019. We find that both dynamical models are consistent with current multi-frequency afterglow observations of the GW170817/GRB170817A event. However, our simulations reveal the possibility of an early rapidly-declining synchrotron emission component, which we refer to as a ``merger flash''. Unlike late afterglow emission, we suggest that rapid follow-up X-ray observations (minutes to hours after the GW event) of the merger flash (including its non-detection) may aid in distinguishing between models.

The details of the numerical setup for both models are presented in Section \ref{sec:setup}. %% and Appendix \ref{Method}
 In Section \ref{sec:sjet}, we demonstrate that successful jets that propagate through, and break out of the NS merger debris cloud naturally develop an angular structure through the interaction with the merger debris. We discuss and analyze the dynamics and the late afterglow radiation of successful structured jets. We present off-axis light curves that match current observations of the afterglow of GRB170817A. In Section \ref{sec:cjet}, we present simulations of the wide engine model and analyze its dynamics and radiative signatures. We then draw comparisons between the narrow engine model and the wide engine model. In Section \ref{sec:multi} we discuss the multiple stages in the computed X-ray light curves.
In Section \ref{sec:early}, we introduce the ``merger flash'', an early rapidly declining light curve component that might be detectable by current or proposed observatories in the minutes following future BNS mergers. A soft X-ray merger flash following GW170817 may have been missed by Swift XRT, due to Earth occultation. For future nearby BNS mergers, the follow-up detection of the merger flash may be possible, particularly at X-ray energies and may help constrain the observer viewing angle. We conclude in Section \ref{sec:conclusion} with a summary of our findings.

\section{Numerical setup} \label{sec:setup}
\subsection{Initial conditions} \label{Method}
Our numerical setup captures the features of BNS mergers that shape the dynamics of the relativistic outflow and its radiative signature. General relativistic magnetohydrodynamics (GRMHD)
simulations of BNS mergers indicate that between $10^{-4}$ and
$10^{-2}$ solar mass ($M_{\odot}$) of neutron star materials are ejected during the coalescence, forming a quasi-spherical debris cloud. The cloud expands mildly relativistically, with typical radial velocity $\sim 0.15 - 0.25 \rm{c}$ (e.g. \citealt{2013PhRvD..87b4001H,2017PhRvD..96l3012S}). The modeling of the ``kilonova'' emission associated with GW170817 reveals that $\sim 10^{-2}\,\rm{M_\odot}$ of neutron rich materials were ejected during the coalescence. We use this cloud mass in our simulations. The ejecta cloud has a slightly oblate geometry and radial stratification; most of its mass is confined in a slow-moving core, while a small amount of mass $\unit[10^{-4}]{M_\odot}$ lies in an extended fast-moving tail,
\begin{eqnarray}
  \rho_c(r,\theta) &=& \rho_c (r/r_c)^{-2}(1/4+\sin^3 \theta) \qquad
  r<r_c \,,\\
    \rho_t(r) &=& \rho_t (r/r_c)^{-n} \qquad r_c<r < 4r_c \,,\\
    v(r) &=& v_cr/r_c \, \qquad r<4r_c\, .
\end{eqnarray}
The density values $\rho_c$ and $\rho_t$ are calculated based on the total mass of the slow-moving core and the fast-moving tail. The density power-law index $n$ is set to 8, the same value adopted in \cite{2017Sci...358.1559K}.
The fast-moving tail is predicted to be the outcome of the shock that forms during the first collision between the merging neutron stars \citep{2014MNRAS.437L...6K,2018arXiv180307595N}. $v_c = \unit[0.2]{c}$ sets the maximal velocity of the core. $r_c=\unit[1.3 \times 10^9]{cm}$ is the core radius. This initial condition is similar to \citealt{2017Sci...358.1559K, 2017arXiv171005896G}. Both the narrow engine model and the wide engine model are evolved in the same merger cloud.

The central engine is initiated when the cloud has evolved for 1 second after the BNS coalescence. We make this choice because it yields GRB prompt emission compatible with the $\unit[1.7]{s}$ time delay between the observation of the GW chirp and the sGRB signal \citep{2017Sci...358.1559K, 2017arXiv171005896G, 2018arXiv180307595N}. The total engine energy is fixed at $\unit[5 \times 10^{50}]{erg}$ (per hemisphere), corresponding to roughly 6\% of the rest mass energy of the merger. Our simulation parameters are summarized in Table \ref{tab:sync}.

% The details of the ejecta cloud profile and the jet engine models are covered in Appendix \ref{Method}. In order to address numerical challenges associated with the Lorentz contraction of ultra-relativistic material, we have implemented a new adaptive mesh refinement prescription into the JET code. This scheme enables us to fully resolve the thin shell of relativistic gas that carries most of the jet's energy.

% An AMR scheme is incorporated into the code to resolve the relativistic thin shell, and has proven to be a robust scheme in terms of accuracy and computational efficiency

% The jet injected through the cylindrical nozzle quickly accelerates and spreads to an opening angle  $\theta_j=1/f\Gamma_0$ where $f=1.4$ \citep{2018MNRAS.473..576G}.

\subsection{Numerical methods} \label{Method}
We conduct 2D axially symmetric relativistic hydrodynamic simulations
with the moving mesh code -- JET \citep{2013ApJ...775...87D}. The jet
engine is injected as a source term for both models. For the narrow
engine model, we choose the jet engine as a nozzle with circular
profile \citep{2015ApJ...813...64D}. For the wide engine model, we
adopt the same injection method of \citealt{2017Sci...358.1559K,
  2017arXiv171005896G}, where a cylindrical nozzle is
used. Throughout, we use an ideal gas equation of state with adiabatic
index 4/3 for simplicity and to allow for comparison with other
  simulations of this event (e.g. \citealt{2018arXiv180307595N}). We
  have performed simulations using the RC equation of state
  \citep{2006ApJS..166..410R} for comparison and find that our results do not qualitatively change.

Our simulations take place on a spherical grid with $N_{\theta} = 160$ zones, evenly distributed in polar angle over the half-sphere. The central engine is modeled by injecting relativistic flow near the cloud center. The inner boundary is located at $4 \times 10^{7}\,\rm{cm}$. The radial grid is logarithmically spaced so that the cell aspect ratio is close to one. Each simulation cell face moves radially with the flow. We adopt an adaptive mesh refinement (AMR) scheme that dynamically refines regions with high Lorentz factor.

\begin{deluxetable}{lcc}[htb!]  
  \tablecolumns{3}
  \tablewidth{2pc}
  \tablecaption{Hydro parameters for the narrow engine model and the wide engine model.\label{tab:sync}}
  \tablehead{
   \colhead {Variable} & \colhead {Narrow Engine}  &
   \colhead {Wide Engine}
  }
  \startdata
  $M_{\rm{cloud}} (DS)$ & $\unit[0.01]{M_{\odot}} $ & $\unit[0.01]{M_{\odot}}$ \\
  $L_{\rm{jet}} (SS)$ & $\unit[2.6\times 10^{50}]{erg\,s^{-1}}$
  & $\unit[2.6\times 10^{50}]{erg\,s^{-1}}$ \\
  $t_{\rm{jet}}$ & $\unit[2]{s}$  & $\unit[2]{s}$\\
  $\Gamma_0$ & 10 & 1.02\\
  $\eta$ & 100  & 20 \\
  $\theta_{\rm{jet}}$ & $0.1$& $0.35$  \\
  $n_{\rm{ism}}$ & $10^{-4},10^{-5}\,\rm{cm}^{-3}$ & $10^{-5}\,\rm{cm}^{-3}$ \\
  \enddata
  \tablecomments{$DS(SS)$ represents double-sided (single-sided). We use the same merger cloud, the same jet engine luminosity $L_{\rm{jet}}$ and engine duration $t_{\rm{jet}}$ for both the narrow engine model and the wide engine model. The cloud mass $M_{\rm{cloud}}$ is $\sim \unit[0.01]{M_{\odot}}$. The initial Lorentz factor $\Gamma_0$, the specific enthalpy $\eta$, and the half opening angle $\theta_{\rm{jet}}$ of the jet engine are set to different values between these two models.}
\end{deluxetable}

\section{Successful jets and the development of angular structure} \label{sec:sjet}
Here we report the dynamics and afterglow signature of a simulation model in which the central engine produces a successful relativistic jet, that is, one that successfully breaks out of the merger cloud and continues propagating into the ISM.

\subsection{Development of the angular structure}
\label{sec:sjet_dynamics}
\begin{figure}[h]
  \centering
  \subfloat[\label{fig:SJet_dynamics_energy}]{
    \includegraphics[clip,width=0.9\columnwidth]
    {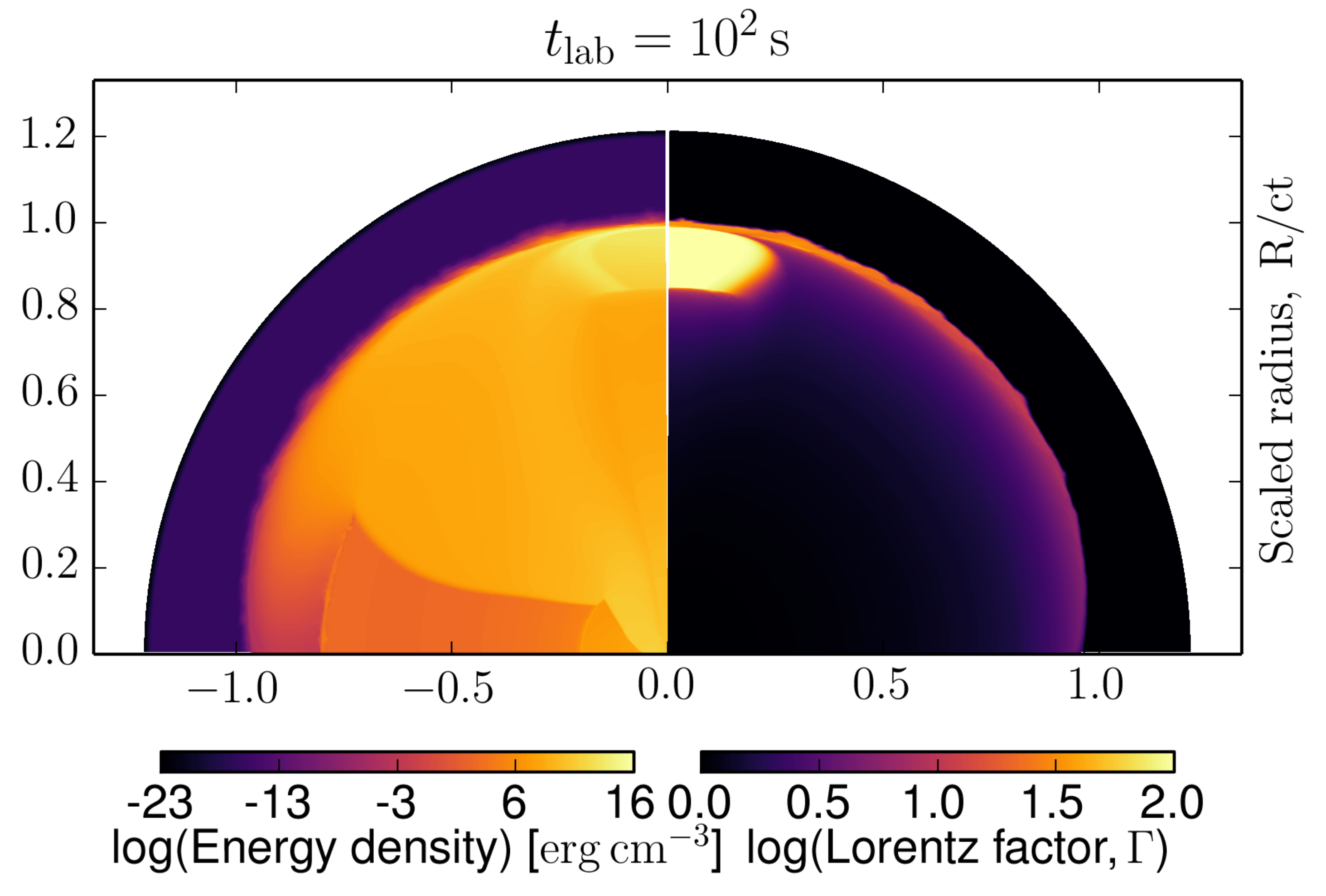}
  }\\[-0.5ex]
  \subfloat[\label{fig:SJet_dynamics_passive}]{
    \includegraphics[clip,width=0.9\columnwidth]
    {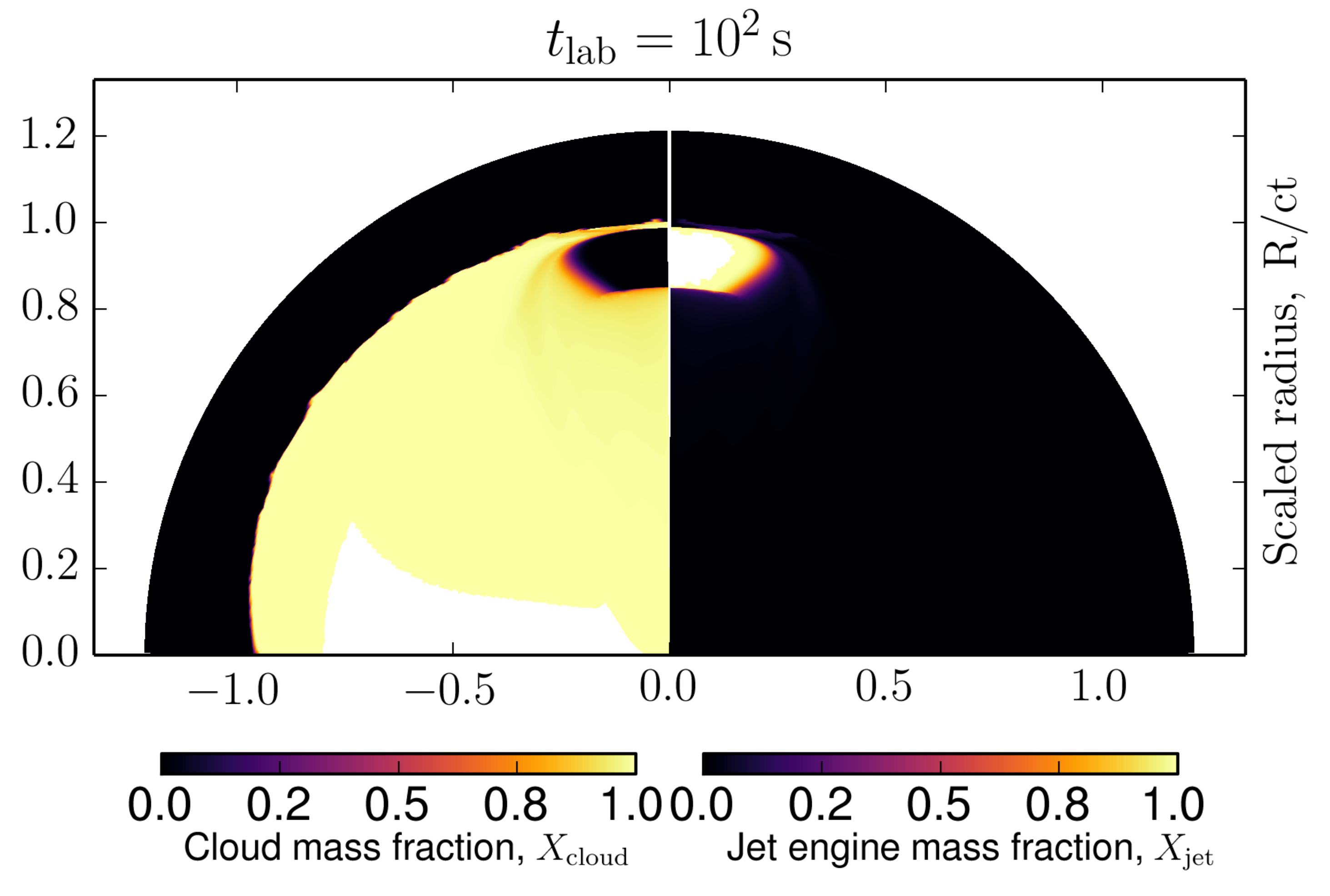}
  }
  \caption{\emph{Narrow engine model} --- (a) The logarithmic total energy density (excluding rest mass; left panel) and the Lorentz factor ($\rm{log}\,\Gamma$, right panel) contour plot taken at the central lab time $t_{\rm lab} = \unit[100]{s}$. (b) The mass fraction contour plot for the merger cloud $X_{\rm cloud}$ (left panel) and the jet engine material $X_{\rm jet}$ (right panel) taken at time $t_{\rm lab} =\unit[100]{s}$. The early domain profile of the narrow engine model contains a still forming  ultra-relativistic core, primarily composed of jet engine materials, surrounded by a mildly relativistic cocoon, composed of merger cloud materials.
    \label{fig:SJet_Contour}}
\end{figure}

In modeling a successful jet, we inject hot, relativistic material within a narrow opening angle (see Table \ref{tab:sync}). The jet drills through the dense core of the merger cloud, and breaks out highly over-pressurized. It drives sideways expansion in the fast-moving, lower density tail of the merger cloud. Eventually, the outflow escapes the cloud altogether, at a radius $\sim \unit[2.4 \times 10^{11}]{cm}$. GRB prompt emission photons are released from the vicinity of this break out radius \citep{2017Sci...358.1559K, 2017arXiv171005896G, 2018arXiv180307595N}. Along the propagation direction, the relativistic GRB ejecta shocks the slower-moving merger debris ahead of it. The internal collision compresses the outflow into a very thin ultra-relativistic core. Meanwhile the rapid lateral expansion of the sideways shock accelerates a mildly relativistic cocoon of neutron star materials, extending to a large lateral angle, as shown in Figure \ref{fig:SJet_Contour}.

\begin{figure}[!ht]
  \centering
  \includegraphics[width=0.5\textwidth]{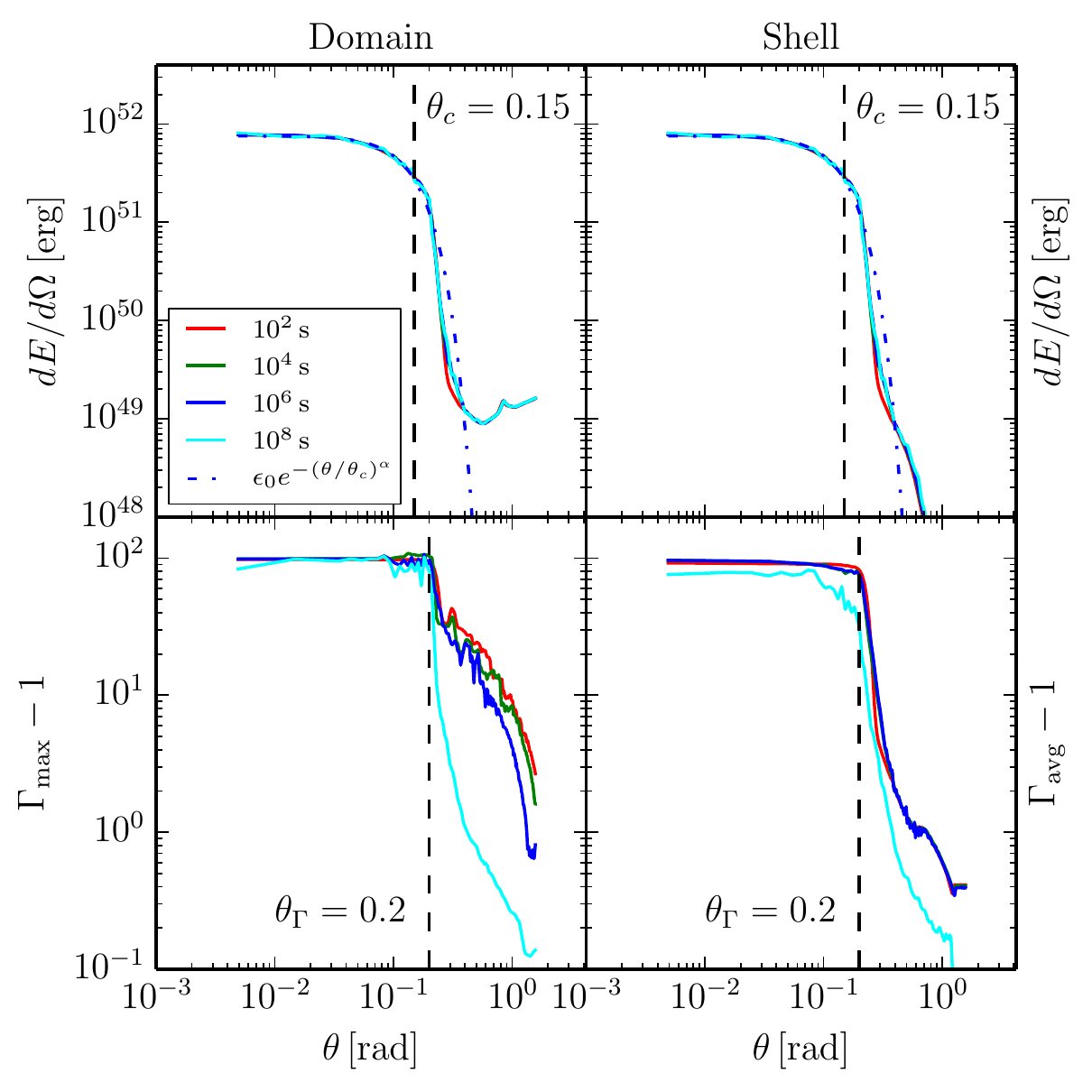}
  \caption{\emph{Narrow engine model} --- Shown in the top row is the
    angular distribution of the total energy (kinetic plus
    thermal). The bottom row shows the maximum Lorentz factor over the
    entire simulation domain (left panel) and the energy-averaged
    Lorentz factor for the relativistic shell (right panel). The
    columns correspond to data taken from the entire domain (left
    column) and from the relativistic shell (right column). Results
    from different time snapshots are colored with $t_{\rm
      lab}=\unit[10^2]{s}$ (red), $\unit[10^4]{s}$ (green),
    $\unit[10^6]{s}$ (blue), and $\unit[10^8]{s}$ (cyan). The angular
    distribution of the total energy is well fitted by a
    quasi-Gaussian model $\epsilon_0 e^{-(\theta/\theta_c)^\alpha}$
    (dot-dashed line) with $4\pi\epsilon_0=\unit[9.6 \times
      10^{52}]{erg}$, $\theta_c=0.15$, and $\alpha=1.93$. The same
    fitting values are found to be appropriate for both the entire
    domain, and the volume occupied by the relativistic shell. The
    half opening angle $\theta_c=0.15$ is shown in vertical dashed,
    purple line. Angles to the left of the vertical line at
    $\theta_\Gamma=0.2$ lie in the relativistic core of the
    jet.} \label{fig:Structure_SJet}  
\end{figure}

After having emerged from the cloud, the jet has developed an angular dependent structure. The angular distribution of the total energy (kinetic plus thermal) is shown qualitatively in Figure \ref{fig:SJet_Contour} and quantitatively in Figure \ref{fig:Structure_SJet}. The jet contains an ultra-relativistic core ($\Gamma\sim 100$), and a mildly relativistic sheath ($\Gamma\sim 10$). In Figure \ref{fig:Structure_SJet}, we differentiate between the relativistic shell ($\Gamma > 1.1$) and the entire domain, labeled as \emph{shell} and \emph{domain}, respectively.

We find that the angular energy distribution ($dE / d\Omega$) for both components is well described by the quasi-Gaussian profile,
\begin{equation}\label{eq:structure}
	dE/d\Omega=\epsilon_0 e^{-(\theta/\theta_c)^\alpha} \, .
\end{equation}
This angular energy distribution is different from the top-hat model typically used in the fitting of GRB afterglow light curves (e.g., \citealt{2012ApJ...749...44V}); it better resembles the model described in \cite{2004ApJ...601L.119Z}. The total energy and the energy-averaged Lorentz factor,
\begin{equation}
  \Gamma_{\rm{avg}} \equiv \frac{\int \Gamma E dV}{\int E dV}\,,\label{eq:gamma_avg}
\end{equation}
of the relativistic shell maintain their initial angular structure for a long period of time $\sim \unit[10^8]{s}$. In Equation \ref{eq:gamma_avg}, $E$ is the local energy density (measured in the lab frame) and $\Gamma$ is the Lorentz factor of the fluid element. The maximum isotropic equivalent energy of the structured jet is $E_{\rm{iso, peak}} = 4 \pi \epsilon_0 \approx \unit[10^{53}]{erg}$. Within an opening angle $\theta_c = 0.15$, the average isotropic equivalent energy of the relativistic core is $E_{\rm iso,avg} = \unit[6 \times 10^{52}]{erg}$, larger than the average isotropic equivalent energy $E_{\rm k, iso} \approx \unit[(1-3) \times 10^{51}]{erg}$, inferred for typical short GRBs, but still within the observed range \citep{2015ApJ...815..102F}.

The angular structure develops as a result of over-pressurized relativistic ejecta escaping the merger cloud into the relatively dilute ambient medium. This results in significant lateral expansion (as depicted in Figure \ref{fig:SJet_Contour}), in addition to radial acceleration. The jet propagating into the ambient medium consists of a shock-heated, baryon-clean core, surrounded by a shock-heated sheath of NS merger ejecta materials.

\subsection{Successful structured jet dynamical evolution}
\label{sec:sjet_dynamics}
\begin{figure}[!ht]
  \centering
  \includegraphics[width=0.5\textwidth]
  {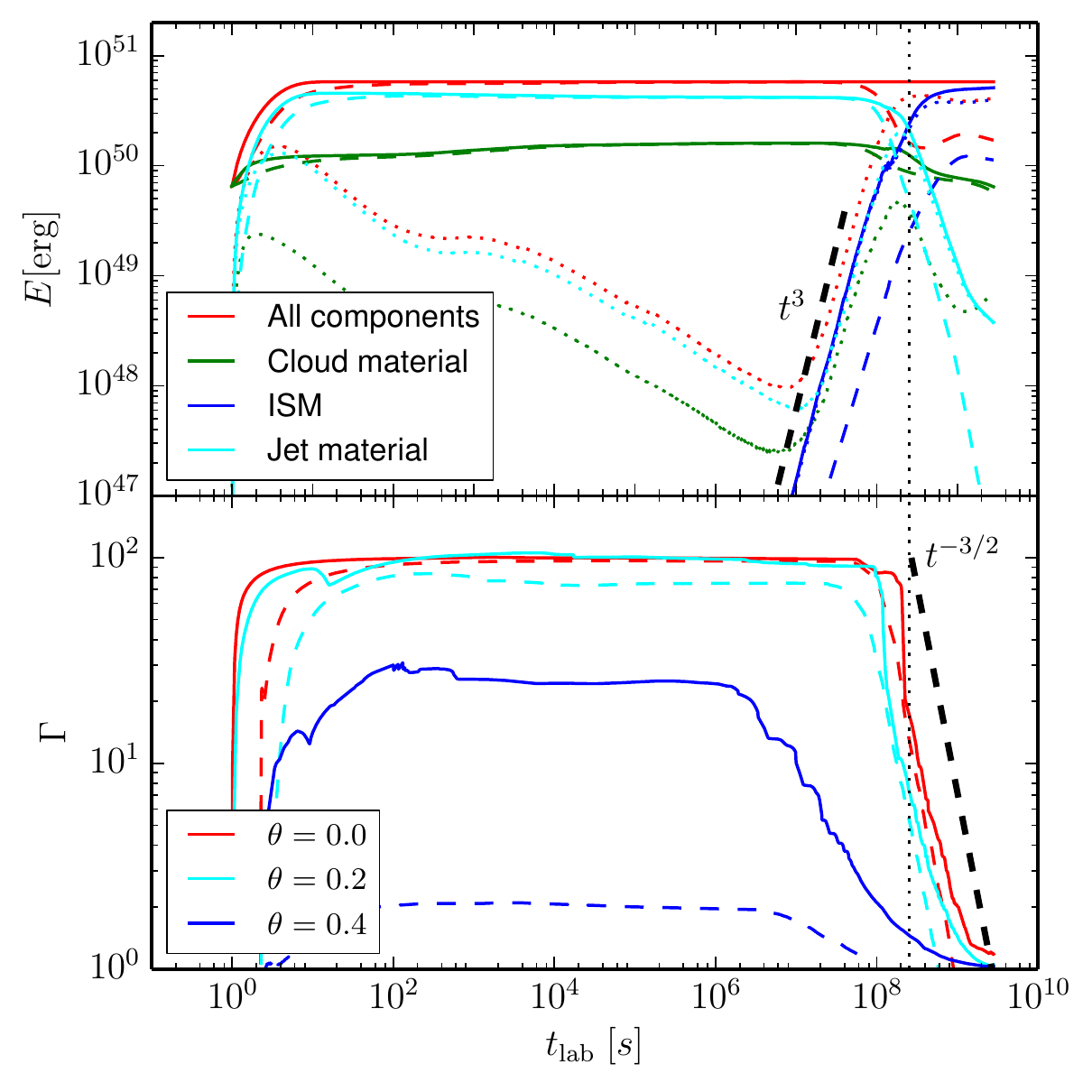}
  \caption{\emph{Narrow engine model} --- The dynamical evolution of various energy components (top panel) and the Lorentz factor (bottom panel), measured in the central lab frame. In the top panel, solid lines represent the total energy (kinetic plus thermal) for four components: everything in the domain (red), the ejecta cloud (green), the ISM (blue), and the jet engine (cyan). Dashed lines (dotted lines) represent kinetic energy (thermal energy) for each component. In less than 100s, most of the thermal  energy converts into the kinetic energy of the cloud and the jet. The residual thermal energy from these two components decreases afterwards in the process of adiabatic cooling. Meanwhile, the energy of the ISM steadily increases. At a later time $t_{\rm lab}>\unit[10^{7}]{s}$, the strong interaction among the jet, the cloud and the accumulated ISM efficiently converts kinetic energy back into thermal energy. The thermal energy of every component increases. In the bottom panel, the time evolution of the maximum Lorentz factor of the relativistic shell along different polar angles are shown in solid lines. The corresponding energy-averaged Lorentz factor are shown in dashed lines. Three dynamical stages are present in the entire life-cycle of the relativistic shell: rapid acceleration, the coasting, and the late deceleration.} 
  \label{fig:Evolution_SJet}
\end{figure}

\begin{figure}[!ht]
  \centering
  \includegraphics[width=0.5\textwidth]
  {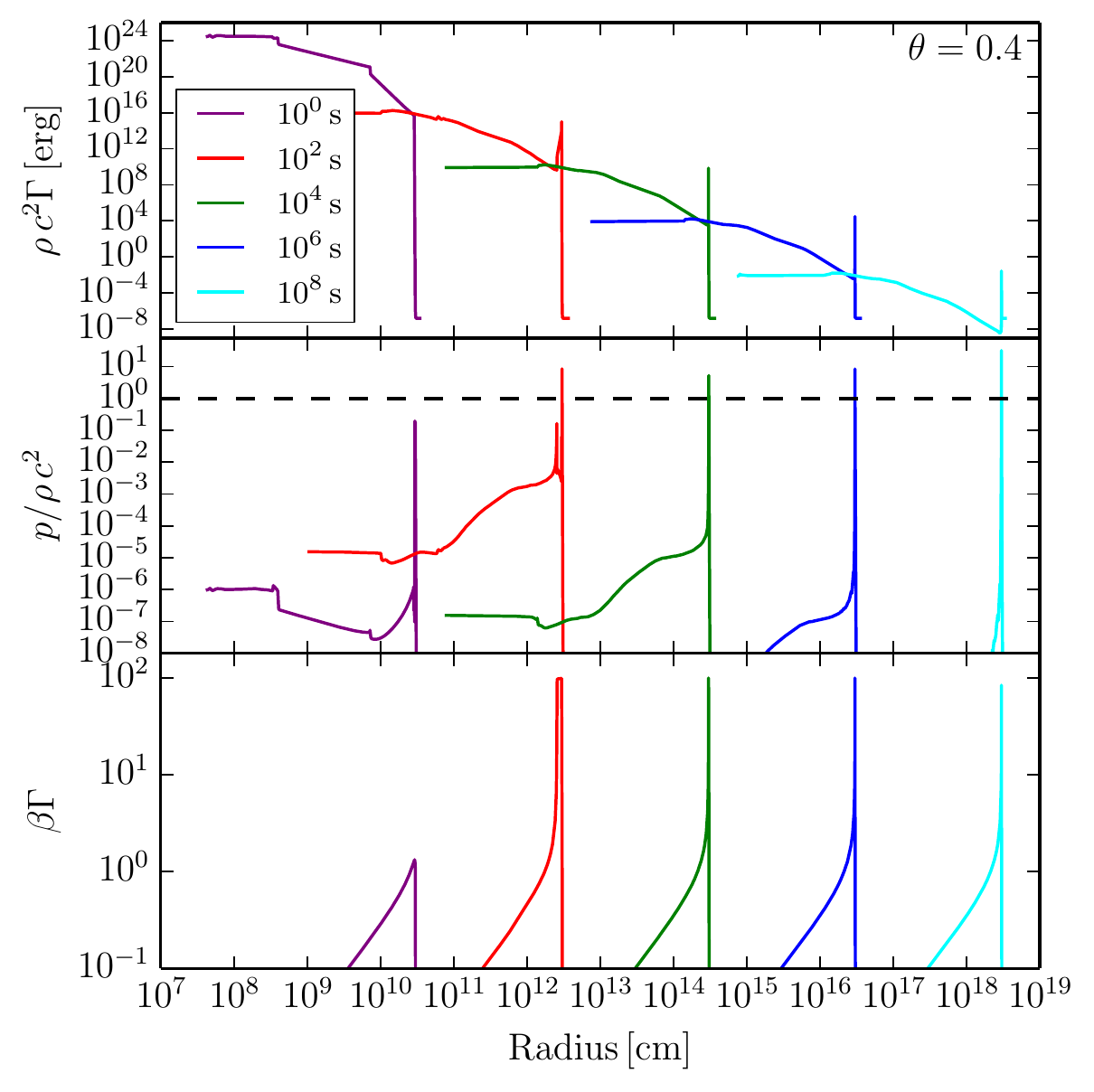}
  \caption{\emph{Narrow engine model} --- Radial profiles of the rest mass
    energy density $\rho c^2 \Gamma$ (top panel),  the ratio of
    comoving pressure and comoving density (middle panel), and
    the four velocity $\beta\Gamma$ (bottom panel) at different time
    snapshots. The radial profiles are taken at the polar angle of
    0.4. Results from different time snapshots are colored with
    $t_{\rm lab}=\unit[1]{s}$ (purple), $\unit[10^2]{s}$ (red), $\unit[10^4]{s}$ (green),
    $\unit[10^6]{s}$ (blue), and $\unit[10^8]{s}$ (cyan). The inner
    and outer boundaries of the simulation domain are moving with the
    outflow of BNS merger ejecta.
    \label{fig:Evolution_Radial}}
\end{figure}

\begin{figure*}[!ht]
  \centering
  \includegraphics[width=0.65\textwidth]{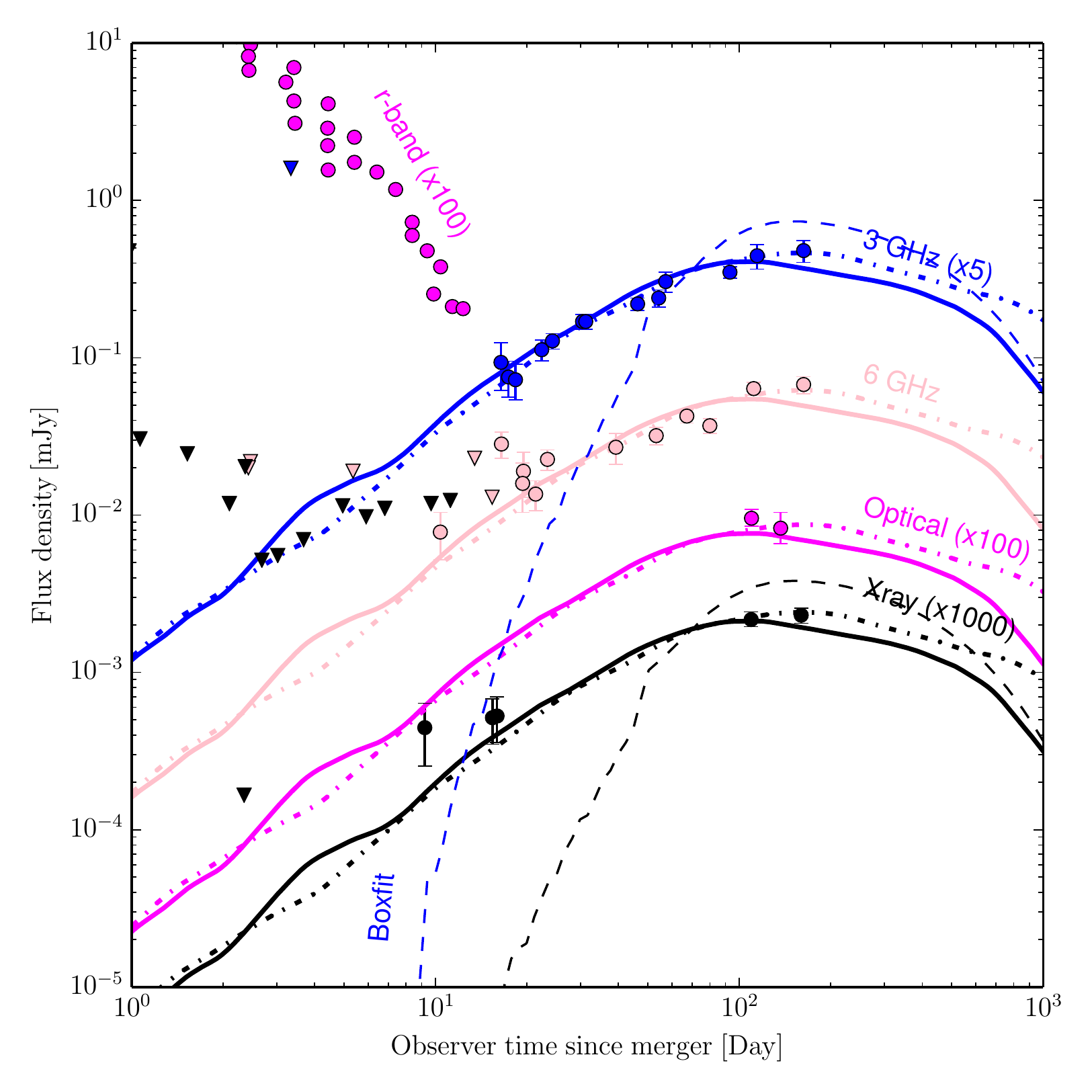}
  \caption{\emph{Narrow engine model} --- Two sets of fitting light curves calculated from numerical simulations of the successful structured jet propagating in two uniform ISM environments. The best-fit radiation parameter values, for the simulation with ISM density $n = \unit[10^{-4}]{cm^{-3}}$, are $\theta_{\rm obs} = 0.34$ ($19.5^{\degree}$), $\epsilon_e = 0.02$, $\epsilon_B = 10^{-3}$, and $p = 2.16$ (solid line). The best-fit radiation parameters for the simulation with ISM density $n = \unit[10^{-5}]{cm^{-3}}$ are $\theta_{\rm obs} = 0.3$ ($17^{\degree}$), $\epsilon_e = 0.1$, $\epsilon_B = 5 \times 10^{-4}$, and $p = 2.16$ (dot-dashed line). These two sets of fitting light curves, smoothed by Savitzky-Golay filter, are consistent with both the early flux upper-limits for non-detection (down triangles) and the detected signals (dots with error bar). The non-detection upper-limits in Radio, X-ray and the non-thermal observations in Radio ($\unit[3]{GHz}$, $\unit[6]{GHz}$), optical ($\unit[5\times 10^{14}]{Hz}$), and X-ray ($\unit[1]{keV}$) are all taken from \cite{2018ApJ...856L..18M}. The early ``kilonova'' r-band data taken from \cite{2018arXiv180102669L} are also presented for comparison. The peak of the light curve is expected to come from the deceleration of the ultra-relativistic core of the successful structured jet. We make a comparison with the results from the top-hat jet model using BOXFIT \citep{2012ApJ...749...44V}. The isotropic equivalent energy and jet half opening angle for the top-hat jet model are set to the representative values of the successful structured jet: $E_{\rm iso, 52} = 6$ and $\theta_j = 0.15$. With the same radiation parameter value $\theta_{\rm obs} = 0.34$, $\epsilon_e = 0.02$, $\epsilon_B = 10^{-3}$, $p = 2.16$, and ISM density $n = \unit[10^{-4}]{cm^{-3}}$, the light curve calculated from BOXFIT (dashed line) rises up faster compared with the light curve from the structured jet model. The peak time and the peak flux of the light curve between these two models are almost the same.}
  \label{fig:spectra_off_axis}
\end{figure*}

After the jet is launched by the central engine, it accelerates by
converting its internal energy into kinetic energy. Over the course of
$\unit[10]{s}$ (as measured in the lab frame, see the bottom panel of Figure \ref{fig:Evolution_SJet}), the jet attains its
terminal Lorentz factor $\eta = 100$  (which is also the specific
internal enthalpy of the engine material at the jet base). During its
propagation through the ejecta cloud, the jet performs work on it. 
In order to determine how the energy is partitioned during this phase of the evolution, we have computed the thermal energy $E_t$ and the kinetic energy $E_k$ for each of three components --- the jet material, the shocked merger cloud (sometimes referred to as ``cocoon'' material), and the ISM. $E_t$ and $E_k$ are given by
\begin{eqnarray}
E_{t,i}& = &\int \left[ \left(e + p \right) \Gamma^2  -p \right] s_i \, dV \nonumber\\
E_{k,i}& = &\int \left[\Gamma (\Gamma - 1) \rho \right] s_i \, dV \, ,
\label{eqn:energy-components}
\end{eqnarray}
where $\rho$, $p$ and $e$ are the co-moving mass density, pressure, and internal energy density, respectively, and $\Gamma$ is the Lorentz factor of the fluid. The subscript $i$ labels the individual component, and the scalar field $s_i$ represents the fraction of each component filling the local volume $dV$; within each cell $\sum_{i=1}^3 s_i = 1$. This decomposition is accomplished by assigning three passive scalars to individual computational cells. The jet material is injected with $s = (1, 0, 0)$, the merger cloud material initially has $s = (0, 1, 0)$, and the ISM has $s = (0, 0, 1)$. As the simulation evolves, individual cells generally acquire some of each component due to mixing at the grid scale. To obtain $E_{k,i}$ and $E_{t,i}$ for each component $i$, we integrate Equations \ref{eqn:energy-components} over the volume.

The top panel of Figure \ref{fig:Evolution_SJet} displays the time evolution of the kinetic and thermal energies in these three components. At the very beginning of jet propagation, the kinetic energy of the jet increases as it accelerates by expending its thermal energy supply. We also observe that simultaneously, the thermal energy content of the cloud material increases. This is the result of $P dV$ work, as well as shock heating, done by the jet on the cloud material as it drills through. Around $\unit[100]{s}$, the jet reaches the outskirts of the merger cloud (see Figure \ref{fig:SJet_Contour}), its kinetic energy saturates and it stops performing work on the cloud material. The jet continues to cool adiabatically (see the dotted lines showing thermal energy in Figure \ref{fig:Evolution_SJet}) as it propagates into the circumburst environment.

The BNS merger event responsible for GW170817 occurred in the
outskirt of an elliptical galaxy
\citep{2017ApJ...848L..22B,2017ApJ...848L..28L}. Low ISM densities are
not unusual in such environments, and therefore we have adopted values in the range $n = \unit[10^{-4} - 10^{-5}]{cm^{-3}}$ for the circumburst number density, which is assumed to be a constant for our discussion. In the co-moving frame of the relativistic shell, the upstream ISM particles stream inward with Lorentz factor $\Gamma$. When an ISM particle crosses the shock front, the direction of its velocity becomes random after multiple collisionless interactions. In the lab frame, the average energy of each downstream ISM proton is $\Gamma^2 m_p c^2$. Detailed studies of jet dynamics and radiation have been covered in GRB reviews (e.g. \citealt{1999PhR...314..575P, 2006RPPh...69.2259M, 2007PhR...442..166N, 2014ARA&A..52...43B, 2015PhR...561....1K}). During the coasting phase of the relativistic jet, its bulk Lorentz factor does not change substantially. However, it performs work on the ISM, while at the same time accumulating mass. The total energy of the swept up ISM is given by
\begin{equation}\label{eq:eism}
    E_{\rm iso,ism} \approx \frac{4\pi}{3}n R^3\Gamma^2\propto t^3\, .
\end{equation}
%
% Throughout the coasting phase, the accumulated work $E_{\rm iso,ism}$ done on the swept-up ISM scales as $t^3$
In Figure \ref{fig:Evolution_SJet}, the energy of the ISM is shown to be increasing throughout the coasting phase $\propto t^3$, in agreement with Equation \ref{eq:eism}. $E_{\rm iso,ism}$ becomes comparable to the energy of the jet at lab time $t_{\rm lab} \sim \unit[2.5\times 10^8]{s}$. This is $\sim 3$ times longer than the predicted deceleration time $R_d/\rm{c}$, according to the estimate of \cite{2015PhR...561....1K},
\begin{eqnarray}\label{eq:dr}
  R_d & = & \left(3E_{\rm iso} / 4 \pi n m_p
  c^2 \Gamma^2 \right)^{1/3} \nonumber \\
  & \approx& 10^{17} E_{\rm iso, 53}^{1/3} n^{-1/3}\Gamma^{-2/3}_{2} \rm{cm} \, ,
\end{eqnarray}
which yields a deceleration time of $\unit[7 \times 10^7]{s}$ for our parameters.

%When the energy of the swept up ISM becomes comparable to the energy $E_{\rm iso}$ of the central engine, %the jet begins to decelerate and $E_{\rm iso, ism}$ deviates from the $t^3$ scaling. 
The jet's transition from the coasting phase to the deceleration phase
is accompanied by the formation of a strong forward shock, which then
propagates into the ISM. A weaker reverse shock, which propagates into
the jet ejecta also forms. The thermal energy of every component increases at the lab time $t_{\rm{lab}}>10^7\rm{s}$. Throughout the deceleration phase, the bulk
Lorentz factor decays as $\Gamma \propto R^{-3/2} \propto t^{-3/2}$
\citep{1976PhFl...19.1130B,1999ApJ...513..669K}. In the bottom panel
of Figure \ref{fig:Evolution_SJet}, the on-axis energy-averaged
Lorentz factor $\Gamma_{\rm{avg}}$ is shown to decay roughly as
$\propto t^{-3/2}$ in agreement with the analytical estimate.

Figure \ref{fig:Evolution_Radial} depicts the evolution of radial
profiles of three physical variables at a representative off-axis polar
angle $\theta=0.4$. After the acceleration phase, a highly relativistic thin shell is
formed. The radial profile of the density follows a power-law decay
with respect to the dynamical radius. At the shock front, the thermal energy density
of the fluid is significant. Both the inner and outer boundaries of the
simulation domain track the radial movement of BNS merger ejecta over
the entire duration of simulations.

\subsection{Successful structured jet afterglow light curve} \label{sec:sjet_rad}
Synchrotron emission using the model of \cite{1998ApJ...497L..17S} can be directly calculated from multi-dimensional hydrodynamical simulation data (e.g. \citealt{2010ApJ...722..235V, 2012ApJ...746..122D}). 
%We do not differentiate between emission from the forward and reverse shocks in this model. However, our model does differentiate between emission that originates from within the GRB ejecta, the shocked cocoon, and the shocked ISM.
%We adopt a standard procedure in GRB afterglow modeling. 
The main parameters determining the synchrotron radiation from the forward shock are the fraction $\epsilon_B$ of post-shock energy residing in magnetic fields, and the fraction $\epsilon_e$ in non-thermal electrons. We further adopt the convention that $\xi_e$ is the fraction of the electrons sharing the electron internal energy $\epsilon_e e$, and that the energy distribution of the relativistic non-thermal electrons is given by $dN/d\gamma \propto \gamma^{-p}$. We assume that $\xi_e = 1$, and the electron spectral index $p$ is taken as a free parameter. We perform simulations of the successful structured jet propagating in low-density environments with two different values for the ISM density, $n = \unit[10^{-4}, 10^{-5}]{cm^{-2}}$. By varying the value of the observer viewing angle $\theta_{\rm obs}$ and the microphysical parameters ($\epsilon_e, \epsilon_B, p$), we obtain two sets of off-axis light curves that match the broadband afterglow observations of GRB170817A.

% The emerged jet has an angular structure almost identical to the analytical Gaussian jet model in Equation \ref{eq:structure} (also adopted in previous studies, e.g. \citealt{2018arXiv180102669L}). Based on our results from Section \ref{sec:sjet_dynamics}, we expect structured jets to be the norm in BNS mergers. Importantly, structured jets produce distinct light curves relative to the uniform top-hat jet models. We have made this comparison in Figure \ref{fig:spectra_off_axis}.

The results of these fits are shown in Figure \ref{fig:spectra_off_axis} (see also \citealt{2018ApJ...856L..18M}). Here we present light curves calculated from the structured jet simulation, contrasted with semi-analytical light curves computed using BOXFIT \citep{2012ApJ...749...44V}, and a simpler top-hat jet profile. The top-hat profile has the same isotropic equivalent energy, $E_{\rm iso,avg} = \unit[6 \times 10^{52}]{erg}$, as the self-consistently simulated jet, and we adopt an opening angle of $\theta_c = 0.15$ which is taken from modeling the simulated jet according to Equation \ref{eq:structure}. Given the same radiation parameters, the light curves calculated from each model peak at roughly the same time, and exhibit similar peak fluxes. However, the early part of the afterglow light curve differs significantly between these two models. In particular, the off-axis light curve from the structured jet brightens earlier than the top-hat jet. The slope of the late decaying light curve from these two models is similar.

The late appearance of the X-ray and radio emission completely rules out any on-axis ultra-relativistic jet models. Indeed, if a relativistic top-hat jet had been pointed away from us, the afterglow emission would have been first detected at a later time, when the emission from the decelerated jet entered our line of sight. The rising light curve from the structured jet is robustly shallower than that of off-axis top-hat models \citep{2018Natur.554..207M}, and is thus detectable at earlier times. The off-axis light curves from the structured jet naturally explain the GRB170817A afterglow emission.

% =====================================================================
% =====================================================================
\section{Wide engine model} \label{sec:cjet}
\begin{figure}[h]
  \centering
  \subfloat[\label{fig:CJet_dynamics_energy}]{
    \includegraphics[clip,width=0.9\columnwidth]{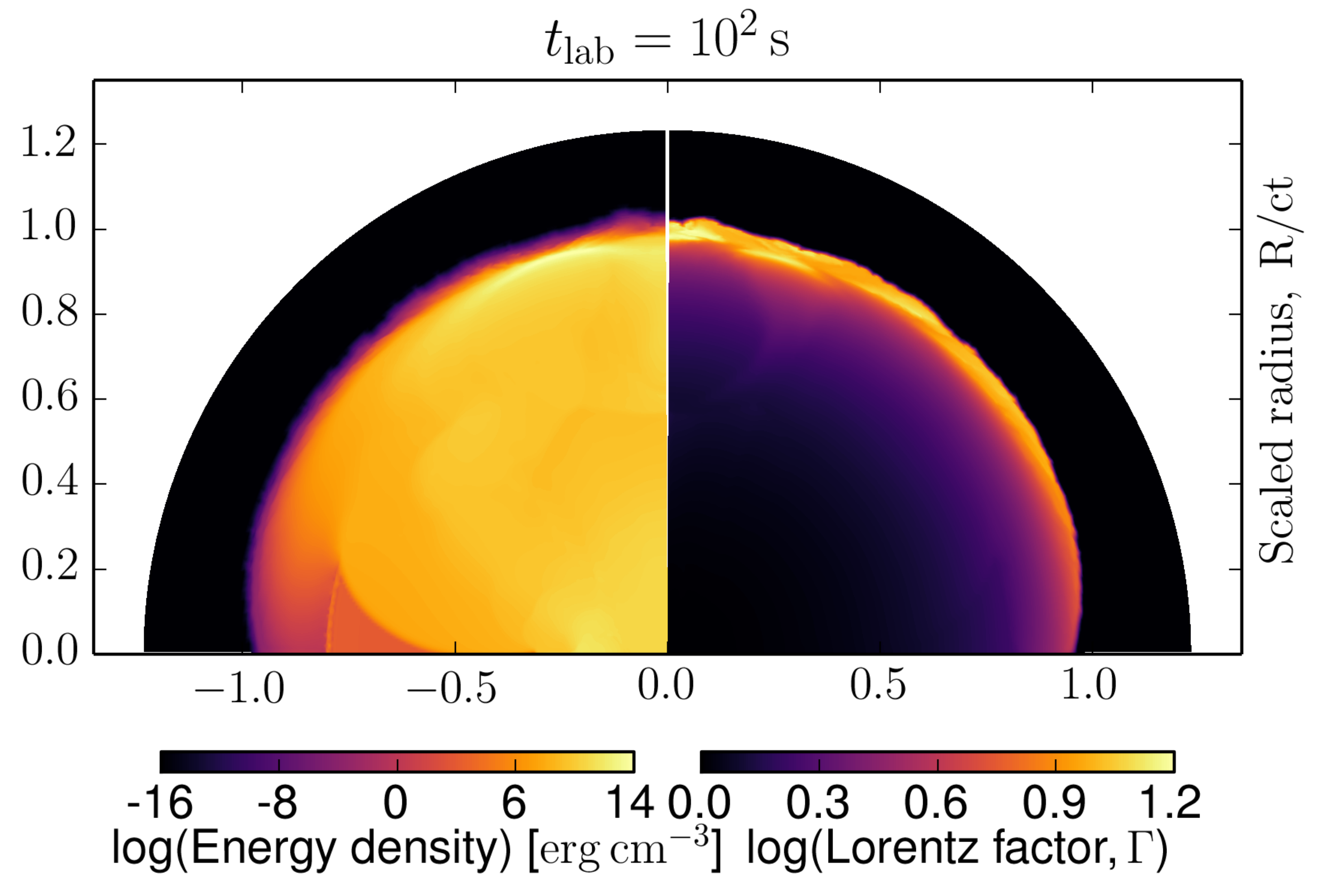}
  }\\[-0.5ex]
  \subfloat[\label{fig:CJet_dynamics_passive}]{
    \includegraphics[clip,width=0.9\columnwidth]{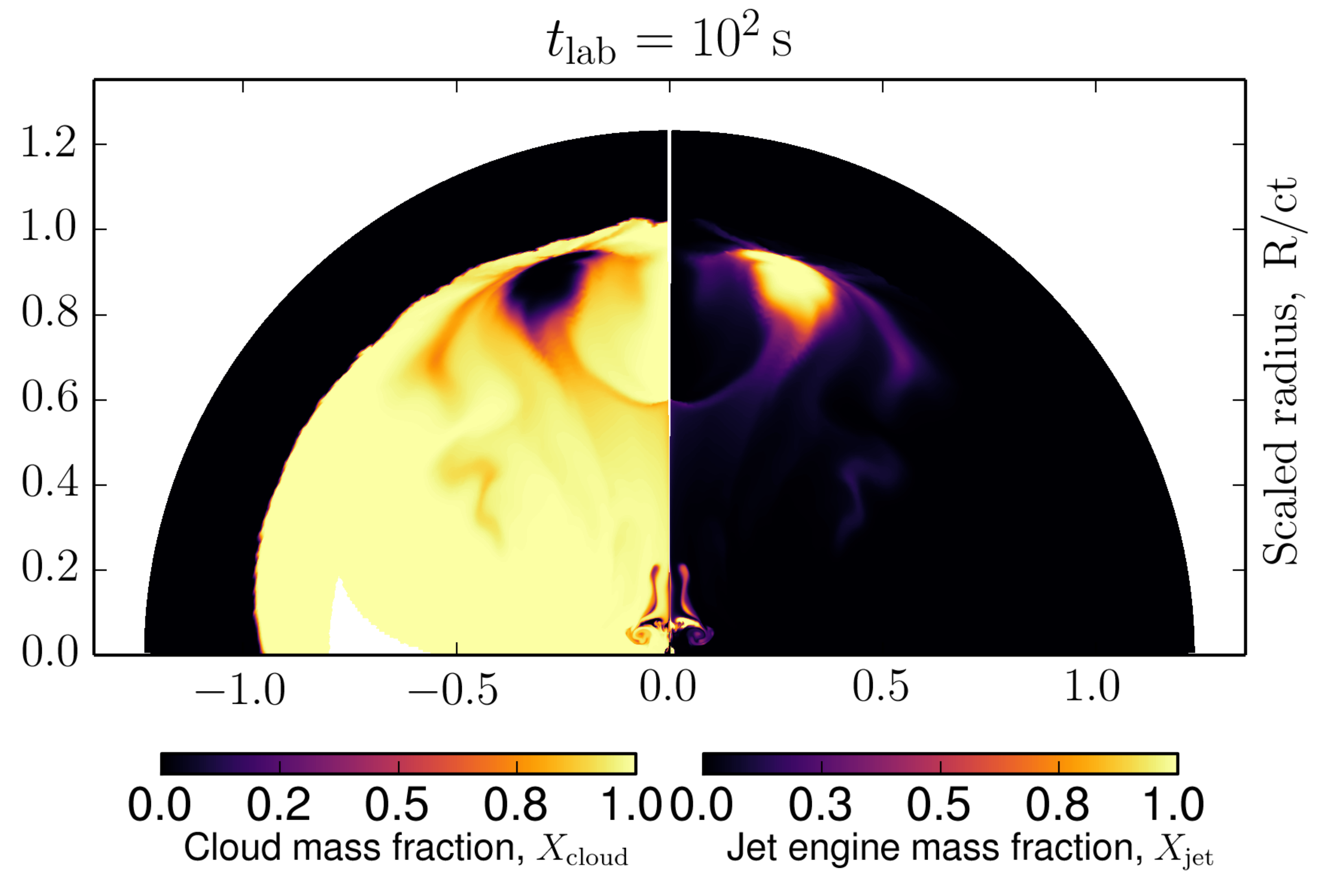}
  }
  \caption{\emph{Wide engine model} --- Same as Figure \ref{fig:SJet_Contour}, but for the wide engine model. The jet engine gets choked by the ejecta cloud, driving a wide spread mildly relativistic shell primarily composed of ejecta cloud materials.
  \label{fig:CJet_Contour}}
\end{figure}

In this section we explore the possibility that the afterglow of GRB170817A was the result of a wide central engine, as may be the case in a ``failed jet'' or ``choked jet'' scenario. A failed jet means that a relativistic outflow was launched by the central engine, but its energy was insufficient for it to emerge well-collimated from the surface of the ejecta cloud.

% =====================================================================
\subsection{Dynamical features}
A wide engine scenario has been invoked in previous studies \citep{2017Sci...358.1559K, 2018MNRAS.473..576G, 2018arXiv180307595N}. During its propagation, the jet is slowed by its interaction with the merger ejecta, and the interaction eventually drives a quasi-spherical mildly relativistic outflow (see Figure \ref{fig:CJet_dynamics_energy}).

In our 2D jet simulations, the relativistic shell
has been well resolved during its propagation inside and outside of
the merger ejecta, via an  adaptive mesh refinement (AMR) scheme.
We find that by resolving the relativistic shell, the
ejecta material that accumulates on top of the jet head is able to
get pushed aside for the narrow engine  model. No strong ``plug'' instability
effect has been observed \citep{2018MNRAS.473..576G,2013ApJ...777..162M,2010ApJ...717..239L}.
For the wide engine model, the jet engine, with a large opening
angle and a small initial momentum, does not have enough power to penetrate the heavy
ejecta material. It diverges halfway through and gets deflected to a large polar angle (see Figure \ref{fig:CJet_dynamics_passive}).

\begin{figure}[!ht]
  \centering
  \includegraphics[width=\columnwidth]{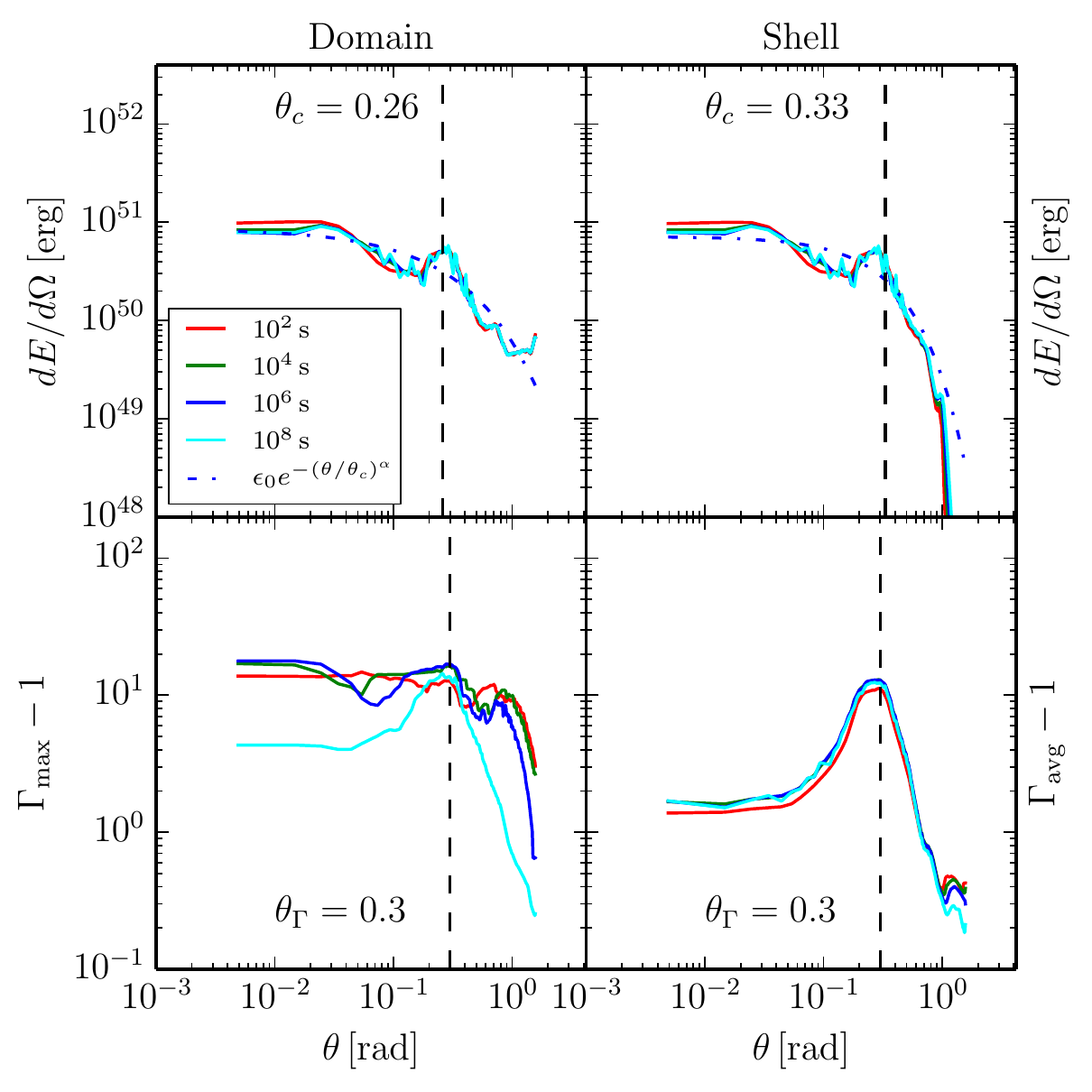}
  \caption{\emph{Wide engine model} --- Same as Figure \ref{fig:Structure_SJet}, but for the wide engine model. Shown in the top row is the angular distribution of total energy (kinetic plus thermal) for the whole simulation domain (left panel) and the relativistic shell (right panel). The bottom row shows the maximum Lorentz factor over the whole simulation domain (left panel) and the energy averaged Lorentz factor for the relativistic shell (right panel). Results from different time snapshots are colored individually with $t_{\rm lab}=\unit[10^2]{s}$ (red), $\unit[10^4]{s}$ (green), $\unit[10^6]{s}$ (blue), $\unit[10^8]{s}$ (cyan). The angular distribution profile of the total energy is well fitted by a quasi-Gaussian model $\epsilon_0 e^{-(\theta/\theta_c)^\alpha}$. We use the fitting value from the relativistic shell: $4 \pi \epsilon_0 \approx \unit[9 \times 10^{51}]{erg}$, $\theta_c = 0.33$, and $\alpha = 1.07$. The wide engine model has a large half opening angle $\theta_c = 0.33$, indicated by the vertical dashed purple line.}
    \label{fig:Structure_CJet}
\end{figure}

An angular structure is also formed in the wide engine scenario, as shown qualitatively in Figure \ref{fig:CJet_Contour} and quantitatively in Figure \ref{fig:Structure_CJet}. The energy angular distribution could be again fitted by a quasi-Gaussian model with an opening angle $\theta_c=0.33$, larger than the opening angle in the narrow engine scenario. Furthermore, the wide jet is found to have a lower peak isotropic equivalent energy $E_{\rm{iso, peak}} = 4 \pi \epsilon_0 \approx \unit[9 \times 10^{51}]{erg}$.

\begin{figure}[!ht]
  \centering
  \includegraphics[width=\columnwidth]{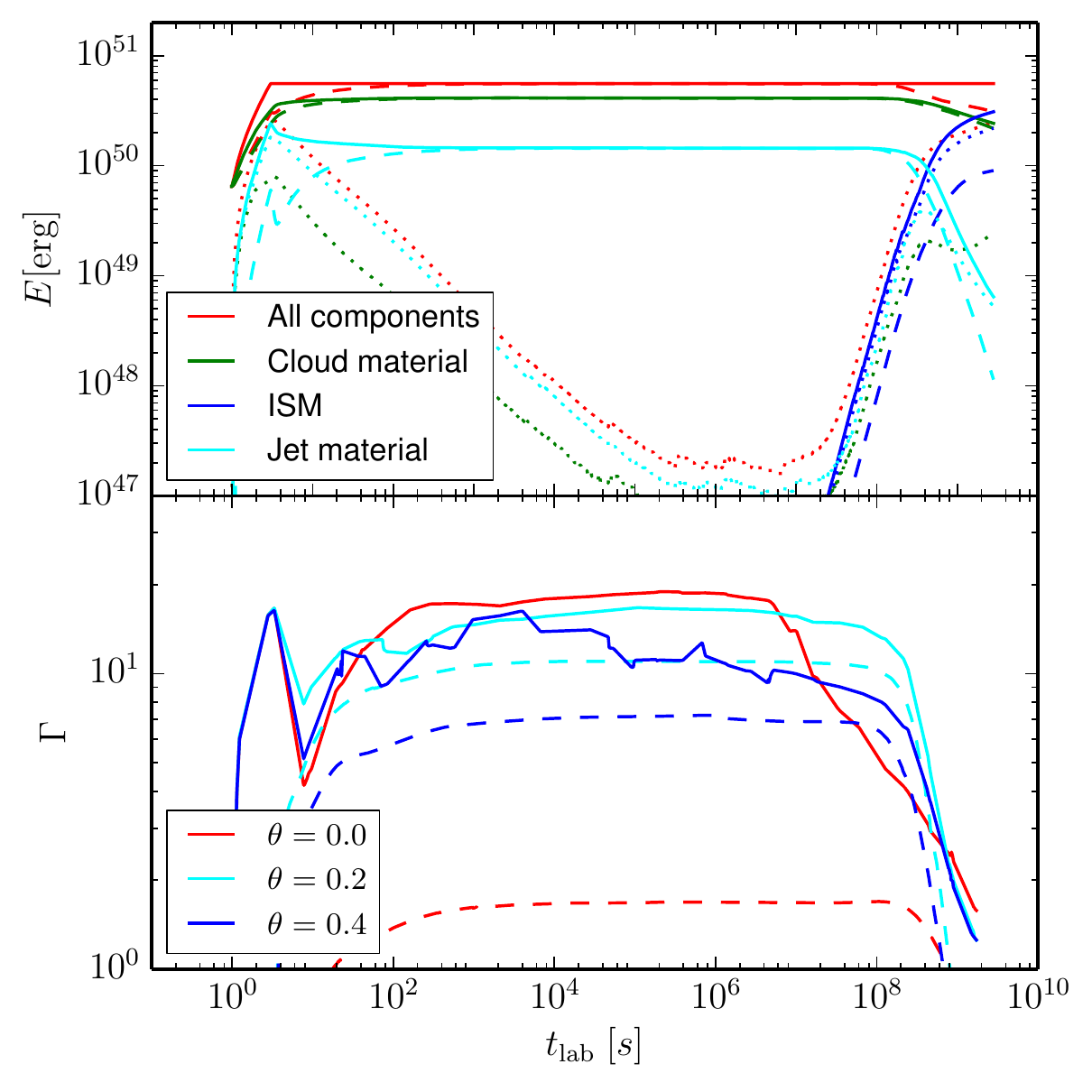}
  \caption{\emph{Wide engine model} --- Same as Figure
    \ref{fig:Evolution_SJet} but for the wide engine simulation: a
    quasi-spherical jet propagating in the ISM environment with
    uniform density $n = \unit[10^{-5}]{cm^{-3}}$. In the top panel,
    solid lines represent the total energy (kinetic plus thermal) for
    four components: everything in the domain (red), the ejecta cloud
    (green), the ISM (blue), and the jet engine (cyan). Dashed lines
    (dotted lines) represent kinetic energy (thermal energy) for each
    component. In less than 10 s, about 80\% of the jet energy is transferred to the cloud. In the bottom panel, the time evolution of the maximum Lorentz factor along different polar angles is shown in solid lines. The energy-averaged Lorentz factor of the relativistic shell is shown in dashed lines. Compared to Figure \ref{fig:Evolution_SJet} for the successful jet case, the Lorentz factor of the relativistic shell is moderate.} \label{fig:Evolution_CJet}
\end{figure}

Whereas in the narrow engine model, roughly 20\% of the jet energy is deposited in the merger cloud, we find that number is  $\sim 80\%$ in the wide engine scenario. This is revealed in the different kinetic energy these two components end to have after acceleration. (see top panel in Figure
  \ref{fig:Evolution_CJet}). The bottom panel in Figure
\ref{fig:Evolution_CJet} shows the time evolution of the energy-averaged
Lorentz factor (Equation \ref{eq:gamma_avg}) as a function of the
polar angle.  $\Gamma_{\rm
  avg}$ ranges from 2 to 10 between polar angles
of 0.0 and 0.4 (also see Figure \ref{fig:Structure_CJet}).

% =====================================================================
\subsection{Afterglow light curve}
\begin{figure*}[!ht]
  \centering
  \includegraphics[width=0.65\textwidth]{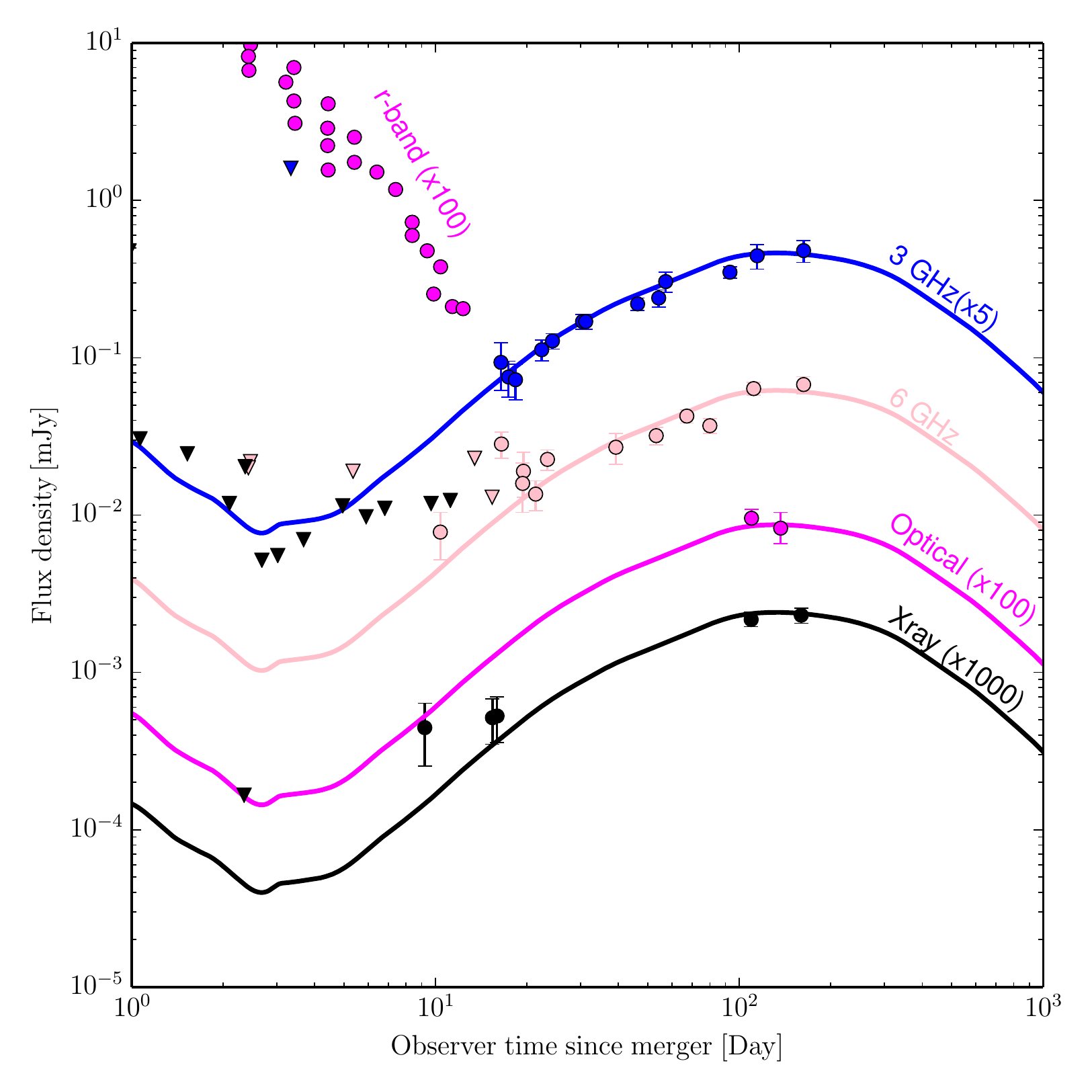}
    \caption{\emph{Wide engine model} --- The fitting on-axis light curve calculated from the numerical simulation of the mildly relativistic quasi-spherical outflow propagating in an ISM environment with uniform density $\unit[10^{-5}]{cm^{-3}}$. The fitting radiation parameter values are $\theta_{\rm obs} = 0$, $\epsilon_e = 0.1$, $\epsilon_B = 2 \times 10^{-3}$, and $p=2.16$. The presented multi-band light curves (smoothed by Savitzky-Golay filter) are consistent with both the early flux upper-limits for  non-detection (down triangle) and detected signals (dots with error bar) in Radio ($\unit[3]{GHz}$, $\unit[6]{GHz}$), Optical ($\unit[5 \times 10^{14}]{Hz}$) and X-ray ($\unit[1]{keV}$). The non-detection upper-limits in Radio, X-ray and the non-thermal observations in Radio ($\unit[3]{GHz}$, $\unit[6]{GHz}$), Optical ($\unit[5\times 10^{14}]{Hz}$) and X-ray ($\unit[1]{keV}$) are all taken from \cite{2018ApJ...856L..18M}. The early ``kilonova'' r-band data taken from \cite{2018arXiv180102669L} are also presented for comparison. We refer the reader to Sections \ref{sec:multi} and \ref{sec:early} for discussions about the early declining light curve.}
    \label{fig:lc_off_axis_CJet}
\end{figure*}

In Figure \ref{fig:lc_off_axis_CJet} we show broad-band afterglow light curves computed from the wide engine model, compared with observations at radio, optical, and X-ray frequencies. For this model, we were only able to obtain a successful fit with a very low external density of $n = \unit[10^{-5}]{cm}^{-3}$.

% =====================================================================
\subsection{Ejecta Lorentz factor distribution}
%
%% \textbf{JZ: I'm not sure about the relevance of this section, and of Figure 8. Also, this material seems to reference successful and unsuccessful scenarios equally, so I'm not sure it belongs in the Wide Engine section.}
%% \textbf{XX: Not quite sure whether the following logic is good or not. This part tries to make comparison between these two models. In the introduction, we also mention we'll compare these two models in the wide engine section.}

\begin{figure}[!ht]
  \centering
  \includegraphics[width=\columnwidth]{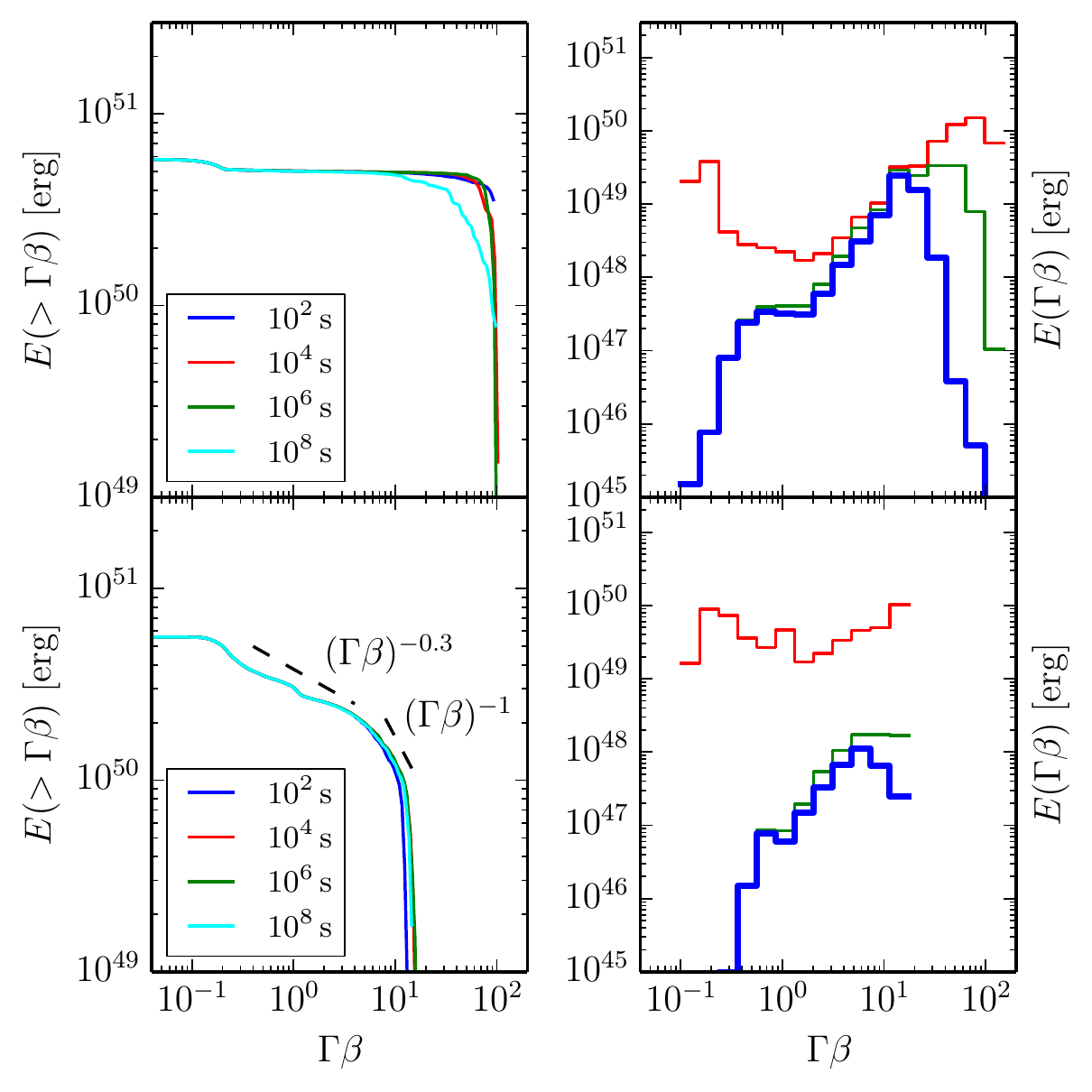}
  \caption{
    The energy four-velocity structure for the narrow engine (top
    row) and the wide engine model (bottom row). The energy four velocity cumulative plot (left column) shows that the outflow from the wide engine model is more stratified radially. In the wide engine model, slow moving materials contain a large fraction of the total energy. In the narrow engine model, most of the energy is contained in the ultra-relativistic shell instead. The right column displays the energy four velocity distribution histogram at $t_{\rm lab}=\unit[10^8]{s}$. During jet's transition from the coasting phase to the deceleration phase, the strong forward and reverse shock generates a large amount of thermal energy in the ISM, also in the entrained ejecta cloud (and the jet engine) materials. The red histogram represents the total energy (kinetic plus thermal) in the entire domain. The green histogram represents the total thermal energy. And the thick blue histogram shows the thermal energy in the ISM only.
  }
   \label{fig:4velocity_CJet}
\end{figure}

In the literature (e.g. \citealt{2018Natur.554..207M}), the stratified
quasi-spherical explosion model utilizes an outflow profile:
$E(>\Gamma\beta)\propto (\Gamma\beta)^{-\alpha}$. The energy power-law
index value $\alpha = 5$ has been found to match early observations
($\unit[<100]{days}$). In the left column of Figure
\ref{fig:4velocity_CJet}, we show the cumulative distribution of energy $E(>\Gamma \beta)$ as a function of four velocity. For the
wide engine model, we find that $E(>\Gamma \beta)$ is not well
characterized by a single power-law. Rather, $\alpha$ increases from
roughly 0.3 on the low velocity end toward 1 at $\Gamma
\beta \sim 10$ (qualitatively similar to results of
\citealt{2018arXiv180300599H}). This is in contrast with the narrow
engine model we explored in Section \ref{sec:sjet}, where a
significant fraction of the energy was seen to reside at high Lorentz
factor.

The right column of Figure \ref{fig:4velocity_CJet} shows the
four velocity distribution histogram of the total energy and the thermal energy at $t_{\rm lab} = \unit[10^{8}]{s}$. A large amount of shock-generated thermal energy resides in the ultra-relativistic shell ($\Gamma > 10$) for the narrow engine  model. The shock-generated thermal energy of the cloud and the jet engine material has higher Lorentz factor compared with the thermal energy of the shock-heated ISM.

% ---------------------------------------------------------------------

% ---------------------------------------------------------------------

% =====================================================================
\subsection{Light curve comparison between narrow and wide engines}
Off-axis light curves from the narrow engine model and on-axis light curves from the wide engine model are able to match the rising light curve observed in the first $\sim 100$ days of GRB170817A. 
The rising light curve component in all of these cases is produced by stratification.   In the case of the narrow engine model, angular stratification is of importance. The high latitude (here defined as near the axis) relativistic material decelerates and adds flux to the rising light curve (see Section \ref{sec:multi}) without producing a sudden brightening, even when the jet core decelerates and comes into view for off-axis observers. If an angularly structured jet is responsible for GRB170817A,  the jet core is probably already been observed.
In contrast, for the wide engine model which produces a quasi-isotropic explosion, the outflow is radially stratified. The slower materials catch up with the decelerating blast wave, driving the rising radiation. In both cases, the light curve comes from the mildly relativistic material (\citealt{2018arXiv180109712N}; see discussion in Section \ref{sec:multi}). Both models predict that the afterglow light curve will decay $\sim 200$ days after the merger, and share roughly the same decay pattern.

% =====================================================================
% =====================================================================
\section{Successful Structured Jet and its Multi-stage light curve} \label{sec:multi}
\begin{figure}[!t]
  \centering
  \subfloat[\label{fig:photo_thermal}]{
    \includegraphics[clip,width=0.9\columnwidth]{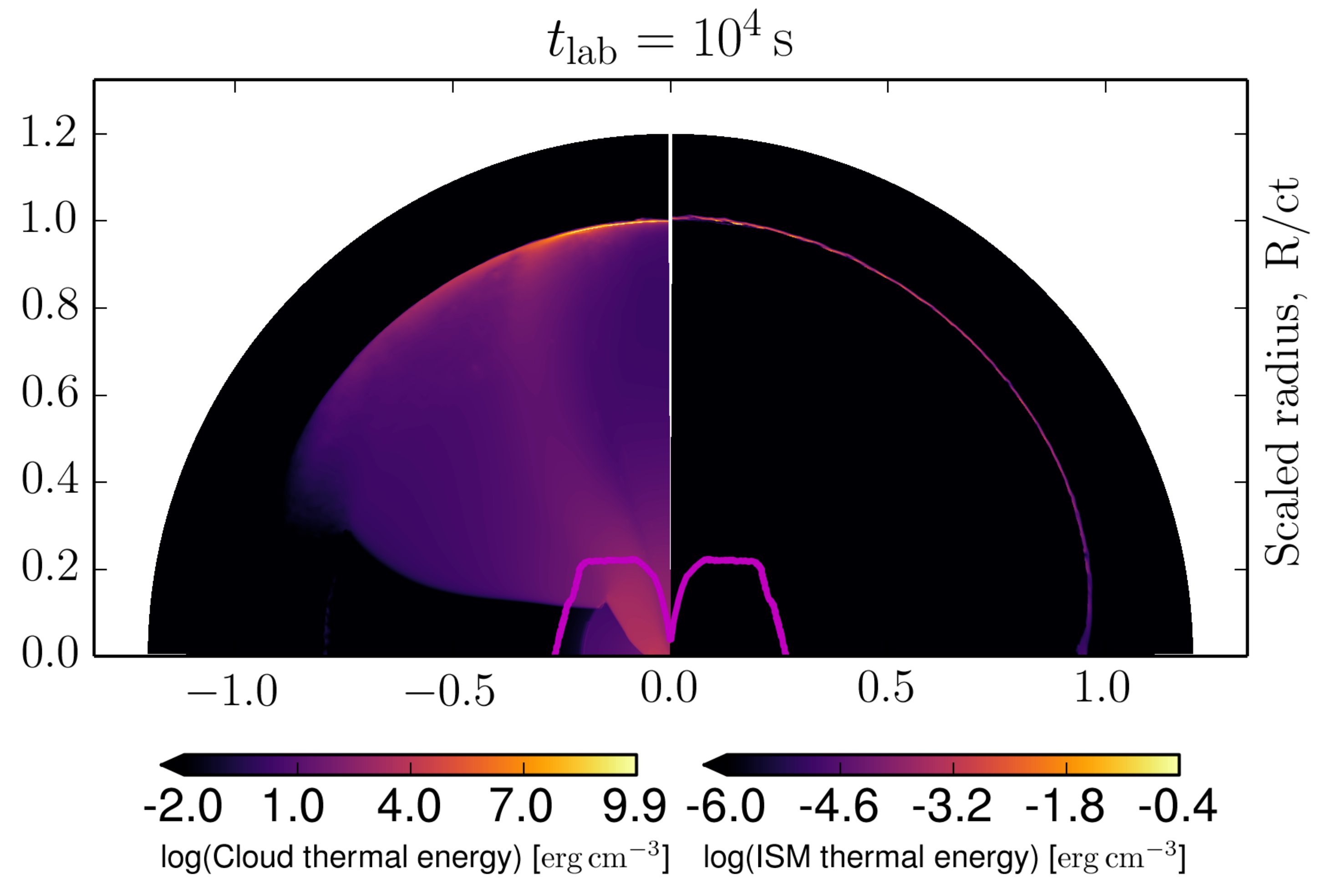}
  }\\[-0.5ex]
  \subfloat[\label{fig:photo_ratio_1e2s}]{
    \includegraphics[clip,width=0.9\columnwidth]{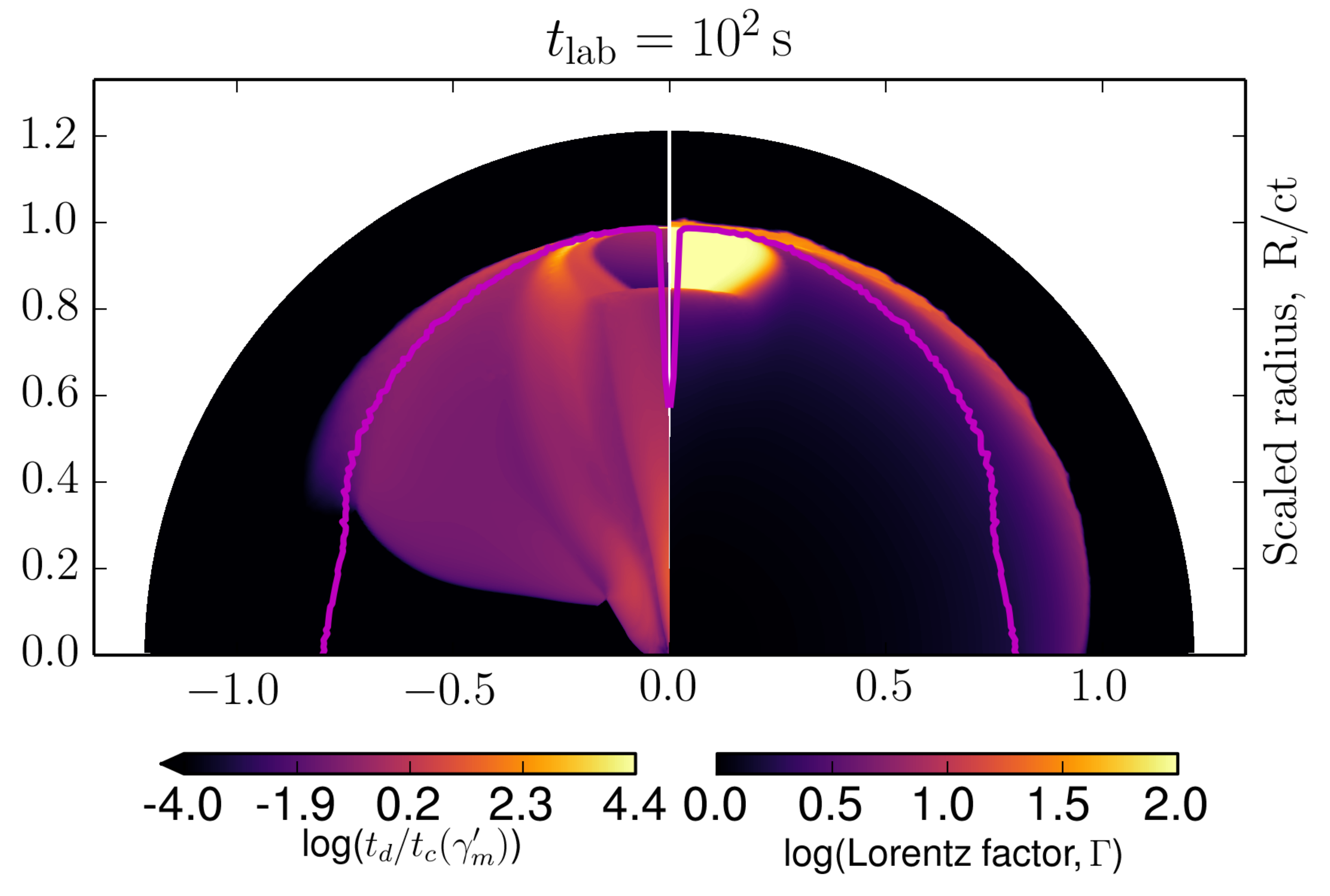}
  }\\[-0.5ex]
  \subfloat[\label{fig:photo_ratio_1e4s}]{
    \includegraphics[clip,width=0.9\columnwidth]{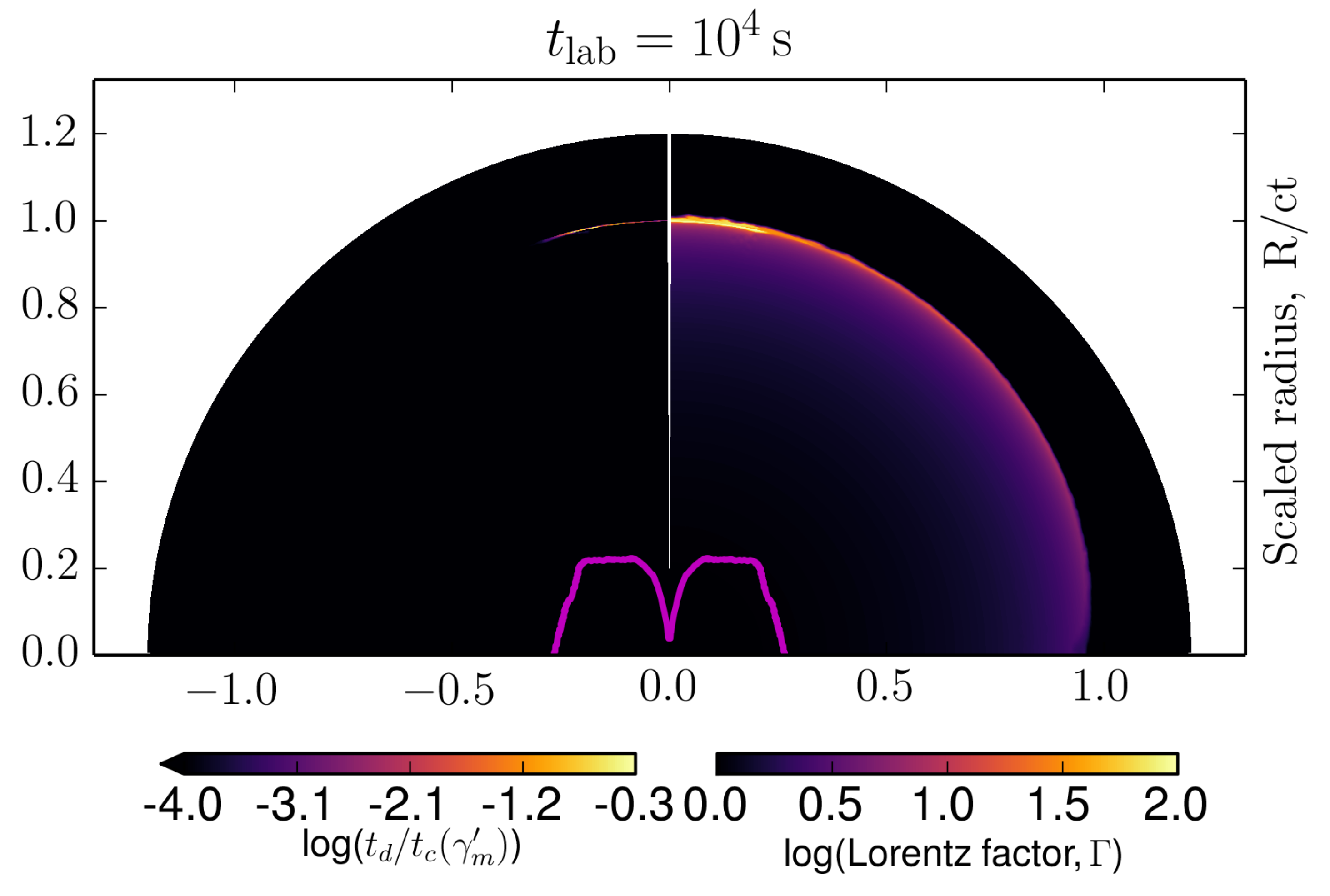}
  }\\[-0.5ex]
  \caption{\emph{Narrow engine model} --- 
    (a) The thermal energy contour plot of the merger ejecta (Left panel) and of the ISM (right panel). The contour snapshot is taken at $t_{\rm lab}=\unit[10^4]{s}$. (b) Left Panel: The ratio contour plot of the dynamical time $t_d$ versus the cooling time $t_c$ at $t_{\rm lab}=\unit[10^2]{s}$. Right Panel: the Lorentz factor contour plot at $t_{\rm lab}=\unit[10^2]{s}$. (c) Same with (b), but at a different time $t_{\rm lab}=\unit[10^4]{s}$. The magenta line indicates the photosphere position viewed by on-axis observers. During the time period $\unit[10^2-10^4]{s}$, the emitting region (thin shell near the shock front) makes the transition from fast cooling to slow cooling.
    \label{fig:photo_contour}}
\end{figure}

\begin{figure*}[!ht]
  \centering
  \includegraphics[width=0.65\textwidth]{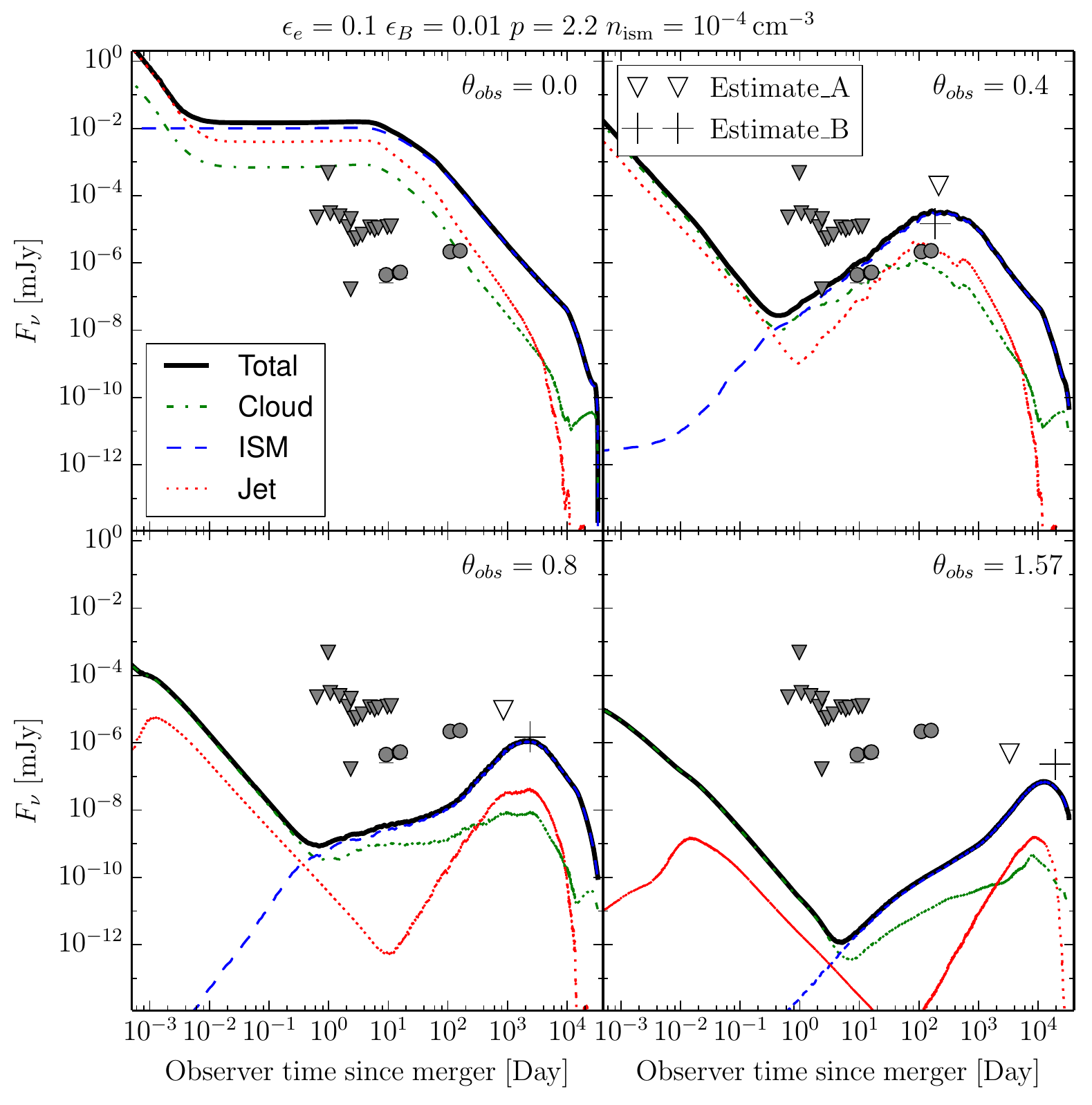}
  \caption{\emph{Narrow engine model} --- The X-ray (\unit[1]{keV}) light curve at different observer angles with radiation parameter $\epsilon_e=0.1,\epsilon_B=0.01,p=2.2$ and ISM density $n=10^{-4}\rm{cm^{-3}}$. Black solid lines represent the total synchrotron emission from the forward jet (radiation from the counter jet is not included). The flux contribution among the three components (the ejecta cloud, the ISM, the jet engine) are separated based on their mass fraction. Synchrotron radiation from the cloud/ism/jet is shown in green dot-dashed/blue dashed/red dotted line, respectively. The on-axis and off-axis light curve displays an universal feature: an initial rapidly declining followed by a late re-brightening. The early declining light curve has internal shock origin (i.e. from the cloud or the jet engine material). The late re-brightening light curve comes from the external shock (i.e. from the shocked ISM). The peak time and the peak flux of the late re-brightening light curve well matches the analytical estimation. Two estimation methods are being used here: Estimate\_A model (hollow down triangle) from \citep{2002ApJ...579..699N,2017arXiv171006421G}, and Estimate\_B model (plus) from \citep{2017MNRAS.472.4953L}. We refer the reader to Section \ref{sec:early} for the viability discussion of the early declining light curve. \label{fig:off_axis_4panel}}
  \end{figure*}

\begin{figure*}[!ht]
  \centering
  \includegraphics[width=0.65\textwidth]{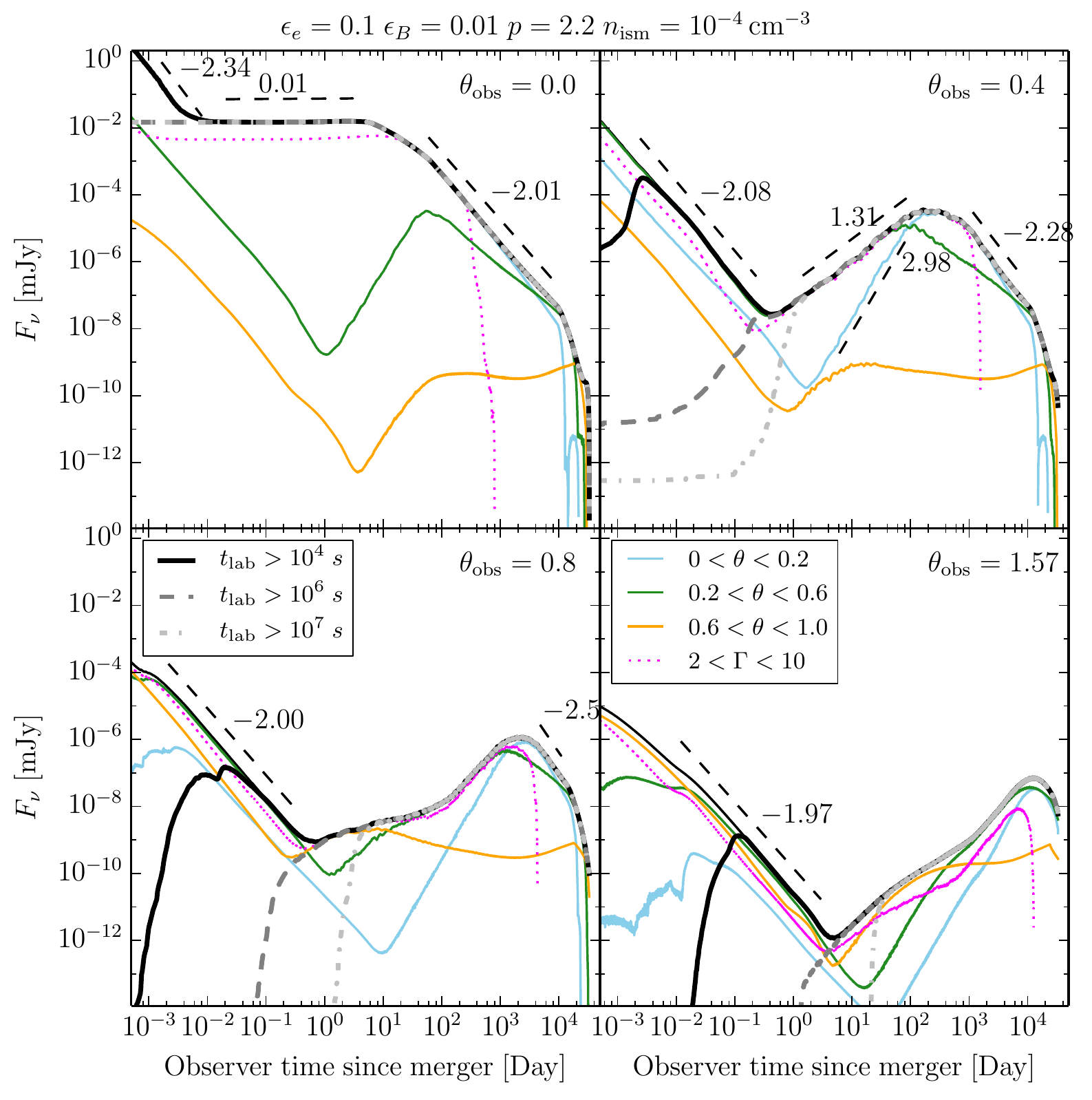}
  \caption{\emph{Narrow engine model} --- The temporal and spatial flux contribution for the on-axis and off-axis X-ray (\unit[1]{keV}) light curve. The post-prompt light curve exhibits three stages: an early afterglow, an intermediate transition, and a late afterglow. The black thick solid lines display the total flux emitted by fluid elements during the time period $t_{\rm lab} > \unit[10^4]{s}$, measured in central lab frame. Gray thick dashed lines (silver thick dot-dashed lines) represent a different time period: $t_{\rm lab} > \unit[10^6]{s}$ ($t_{\rm lab}>\unit[10^7]{s}$). The early afterglow part is emitted before $t_{\rm lab} = \unit[10^{6}]{s}$. Both the intermediate transition and the late afterglow come from a time period $t_{\rm lab}>\unit[10^{6}]{s}$ (see also Figure \ref{fig:Evolution_SJet}). The synchrotron emission from different angular regions in the domain is shown in thin solid lines. The blue/green/orange line shows the flux contributed by fluid elements within a domain lateral angle extending from $0.0/0.2/0.6$ to $\unit[0.2/0.6/1.0]{[rad]}$ (flux from the fluid with domain angle larger than \unit[1.0]{rad} is minimal and not presented here). The early afterglow and the intermediate transition light curve for off-axis observers initially comes from the angular region closer to the line of sight. The late afterglow comes from the central angular region $0<\theta < 0.2$. The magenta dotted line displays the flux contributed by fluid elements with Lorentz factor larger than 2, but smaller than 10. All of the observed emission before the observer time $t_{\rm{obs}}\sim 200$ days originates from the relativistic thin shell with Lorentz factor larger than 2 (except for $\theta_{\rm obs} = 1.57$). Part of the late afterglow comes from the decelerated sub-relativistic materials.
    \label{fig:off_axis_4panel_temporal_spatial}}
\end{figure*}

Section \ref{sec:sjet} presented the dynamics and afterglow radiative sigatures from the successful structured jet simulations. Here we analyze the radiative features in detail, focussing on the X-ray light curve. In order to post-process each simulation output in the time series of saved data files and compute synchrotron light curves, we first estimate the photosphere location of the ejecta outflow by integrating the optical depth along the observer's line of sight:
\begin{equation}
  \tau = \int^{\infty}_{r_{ph}}\sigma_T \Gamma(1-\beta
  \cos\theta)n^{\prime} dl \,,\label{eq:photosphere}
\end{equation}
where $\beta$ is the absolute value of the velocity normalized by the speed of light, $\theta$ is the angle between the velocity vector and the observer's line of sight, $dl$ is the distance along line of sight, $n^{\prime}$ is the proper electron number density, and $\sigma_T$ is the Thomson cross section for electron scattering \citep{2011ApJ...732...26M}. The photosphere position, corresponding to the $\tau\sim 1$ surface, is used to identify optically thin regions of the simulation volume. The photosphere position for on-axis observers is shown in Figure \ref{fig:photo_contour}. We calculate the synchrotron emissivity from simulation cells above the photosphere to compute light curves. 

To determine whether the electrons in a fluid element are in the fast cooling or slow cooling regime, we calculate the dynamical time $t_d$, the minimum Lorentz factor $\gamma_m^{\prime}$ of the electrons, and the associated cooling time $t_c(\gamma_m^{\prime})$, according to:
\begin{eqnarray}
	t_d&=& R/(c\Gamma^2)\, \\	
  \gamma_m^{\prime} &=& \frac{(p-2)/(p-1) \epsilon_e
    e^{\prime}}{(\rho^{\prime}/m_p)m_e c^2}\, ,\\
  t_c(\gamma^{\prime})&=&3m_e c/(4\sigma_T\Gamma\gamma^{\prime} \epsilon_B
  e^{\prime})\,.
\end{eqnarray}
When the dynamical time exceeds the cooling time, $t_d>t_c(\gamma_m^{\prime})$, the fluid element is in the fast cooling regime. At $t_{\rm lab}>\unit[10^2]{s}$, the fast cooling regions fall behind the photosphere  and are thus not included in the synchrotron radiation calculation. By $t_{\rm lab}=\unit[10^4]{s}$, the entire simulation volume is in the slow cooling regime \footnote{In the radiation calculation we include the effect of electron cooling using a global estimate where the electron cooling time equals the lab frame time since the BNS merger.} (see Figure \ref{fig:photo_ratio_1e2s} -\ref{fig:photo_ratio_1e4s}).

\subsection{X-ray light curve and the comparison with analytic estimates} \label{sec:lcan}
In Figure \ref{fig:off_axis_4panel} we display the X-ray synchrotron
emission light curve calculated from the narrow engine simulation with ISM density $n=10^{-4}\,\rm{cm^{-3}}$. The
light curve covers seven orders of magnitude in observer time starting
from one minute and extending to $\sim \unit[30]{years}$ after the BNS
merger. The microphysical parameters, $\epsilon_e$ (the relativistic
electron fraction), $\epsilon_B$ (the magnetic energy fraction), and
$p$ (the slope of the electron distribution), are set to standard
values: $\epsilon_e=0.1,\epsilon_B=0.01,\,\rm{and}\,p=2.2$. In order
to check the accuracy of our numerical light curves, we compare the
peak time and peak flux to estimates from existing analytical
models. The first model (Estimate\_A) is based on an adiabatic double-sided top hat jet with total kinetic energy $E_k$, an initial opening angle $\theta_j$ and a simple hydrodynamical evolution model \citep{2017arXiv171006421G,2002ApJ...579..699N}. The peak time of the off-axis afterglow light curve occurs when the bulk Lorentz factor of the top hat jet drops to $\Gamma=1/\theta_{\rm{obs}}$. The peak time and the peak flux are given as
\begin{eqnarray}
  t_{\rm{peak}}(\theta_{\rm{obs}}) &=&
  0.7(1+z)\left(\frac{E_{\rm{k,51}}}{n_0}\right)^{1/3}\left(\frac{\theta_{\rm{obs}}}{0.1}\right)^2\ \rm{days}\,
  ,\\
  F_{\nu_m < \nu < \nu_c}^{\rm{peak}}(t) &=& 0.6 \frac{g_1(p)}{g_1(2.2)}(1+z)^{(3-p)/2} D_{L28}^{-2}\\
  &\times&  \epsilon_{e,-1}^{p-1}\epsilon_{B,-2}^{\frac{p+1}{4}}n_0^{\frac{p+1}{4}}E_{\rm{k,50.7}}
 \nu_{14.7}^{(1-p)/2}\theta_{obs,-1}^{-2p}\ \rm{mJy}\nonumber\,.
\end{eqnarray}
In another model (Estimate\_B), the projected surface area and the solid-angle of emission are taken into consideration \citep{2017MNRAS.472.4953L}. The peak time and the peak flux are given by
\begin{eqnarray}
  t_{\rm{peak}}(\theta_{\rm{obs}}) &=&
  195\left[\frac{(5+p)(7-p)^{1/3}}{(p-1)^{4/3}}\right]\left(\theta_{\rm{obs}}-\theta_j\right)^{8/3}\,
  ,\\
  &\times& n_{-1}^{-1/3}E_{\rm{k,52}}^{1/3}\ \rm{days}\nonumber\, ,\\
  f_{\nu_m <\nu<\nu_c}^{\rm{peak}}&=&
  C(p)f(\theta_{\rm{obs}},\theta_j)\left(\theta_{\rm{obs}}-\theta_j\right)^{2(1-p)}\nu^{(1-p)/2}\\
    &\times& E_k\ n^{(1+p)/4}\epsilon_B^{(1+p)/4}\epsilon_e^{p-1}
  D_L^{-2} \rm{erg\ s^{-1} cm^{-2} Hz^{-1}}\nonumber\, ,
\end{eqnarray}
where the expressions for $g_1(p), C(p),f(\theta_{\rm{obs}},\theta_j)$ are given in \cite{2017arXiv171006421G} and \cite{2017MNRAS.472.4953L} and $E_k=E_{\rm{iso}}\theta_{j}^2/2$ is the jet kinetic energy (double sided). Here we model the ultra-relativistic core of the structured jet simply as a uniform top-hat jet with kinetic isotropic equivalent energy $E_{\rm{iso,52}} \sim 6$, and jet half opening angle $\theta_j\sim 0.15$. The ISM density is set to the value adopted in the simulation, $n= 10^{-4}\,\rm{cm}^{-3}$. As shown in Figure \ref{fig:off_axis_4panel}, the analytical estimate of the peak time and the peak flux at different viewing angles from Estimate\_B is in agreement with the calculated light curve.

\subsection{X-ray light curve shape}
The on-axis X-ray light curve shown in Figure \ref{fig:off_axis_4panel} (top left panel) displays three temporal power-law segments: 1) an early time steep decay phase $F_{\nu}\propto t^{-\alpha} $, with temporal index $\alpha_1\sim 2.3$. 2) a shallow decay (plateau) phase with index $\alpha_2\sim 0$, and 3) a later decay phase with index $\alpha_3\sim 2$. These light curve components share similarities with the on-axis X-ray light curves for GRBs observed by Swift \citep{2006ApJ...642..354Z,2015PhR...561....1K}.

The off-axis light curves shown in Figure \ref{fig:off_axis_4panel} (top right and bottom two panels) exhibit an early rapidly-fading phase followed by a later re-brightening. Both on-axis and off-axis light curves have three common stages: an early declining afterglow, an intermediate transition phase (the rising part), and a late afterglow (the late declining part). The early declining emission mainly comes from the shock-heated cloud (for the on-axis light curve, it is the jet instead) and decays on a time scale of minutes to days depending on the viewing angle. The flux contribution from the external shock in the ISM steadily increases. Both the intermediate transition phase and the late afterglow light curve come from the shock-heated ISM (see Figure \ref{fig:off_axis_4panel}).

\subsection{Temporal decomposition of the light curve}
Figure \ref{fig:off_axis_4panel_temporal_spatial} shows the temporal and spatial decomposition of the computed light curve. First, we separate the entire simulation duration into three lab time periods: $t_{\rm lab} > 10^4$s; $t_{\rm lab} > 10^6$s; $t_{\rm lab} > 10^7$s. Most of the early declining flux observed during the first $\sim$day of observer time is emitted during the lab frame time period $\unit[10^4]{s}<t_{\rm{lab}}<\unit[10^6]{s}$ for both on-axis and off-axis light curves. As seen in Figure \ref{fig:Evolution_SJet}, before lab time $t_{\rm lab} \sim \unit[10^7]{s}$, almost all of the thermal energy in the domain is in the jet engine and the merger cloud material. The early declining emission is due to the cooling of the post-shock jet engine and merger cloud material. At later times, the relativistic shell experiences strong forward and reverse shocks as it sweeps up and shocks the ISM. The internal energy from the shock-heated ISM then begins to play an important role in the synchrotron emission. The flux during the intermediate transition and the late afterglow is mainly emitted during the lab time period $t_{\rm lab}>10^{7}\rm{s}$, consistent with the dynamical evolution of the thermal energy. The turning point between the early declining and the intermediate transition phases depends on the Lorentz factor of the emitting shell. For on-axis observers, photons radiated at lab frame time $t_{\rm lab}$ and lab frame position $\boldsymbol{r}$ will reach an observer at observer time $t_{\rm obs}$ (see e.g. \citealt{1999PhR...314..575P, 2006RPPh...69.2259M})
\begin{equation}\label{eq:obs_time}
  t_{\rm{obs}} = (1 + z) (t_{\rm lab} - \boldsymbol{r} \cdot \boldsymbol{n}/c) \sim (1+z)t_{\rm lab} / 2\Gamma^2 \, .
\end{equation}
where $\boldsymbol{n}$ is a unit vector pointing in the direction toward the observer.
Along the jet propagation direction, the  bulk  Lorentz factor of the on-axis relativistic shell is $\sim 100$. A photon emitted from the shell at lab time $t_{\rm lab}\sim\unit[10^7]{s}$  will thus be received by on-axis observers at observer time $t_{\rm obs} \sim \unit[8]{min}$. This roughly determines the turning point between the initial steep decay and the shallow decay phase of the on-axis light curve shown in Figures \ref{fig:off_axis_4panel} and \ref{fig:off_axis_4panel_temporal_spatial}.

Previous studies suggest that the initial steep decay phase of the on-axis light curve is linked to the tail of the GRB prompt emission, and has internal shock origin (e.g. \citealt{2005ApJ...635L.133B,2015ApJ...806..205D}). Our result supports this interpretation.  We refer the reader to Section \ref{sec:early} for further discussion about the early declining light curve. The shallow decay phase has been previously interpreted in the context of a refreshed shock model \citep{1998ApJ...496L...1R,2000ApJ...535L..33S,2006ApJ...642..354Z}. Based on our simulation results, we find the duration of the shallow decay phase depends on the initial bulk Lorentz factor and the isotropic equivalent energy of the relativistic jet. It also depends on the ambient density. The typical time scale of the plateau phase observed for classical GRBs is $10^3\sim10^4\ s$ \citep{2015PhR...561....1K}. The BNS case considered here, an energetic jet propagating in a very low density environment, results in a duration longer than this.

\subsection{Angular decomposition of the light curve}
For the off-axis light curve, the early rapidly-fading and later re-brightening behavior distinguishes it from the on-axis light curve. In Figure \ref{fig:off_axis_4panel_temporal_spatial}, we divide the simulation domain into angular regions and calculate the flux contribution from each one of them. Off-axis observers will first detect radiation from the part of the outflow that is moving toward them, i.e. in the direction of the observer's line of sight. As time goes on, the decelerated relativistic shell at higher latitudes (i.e. closer to the polar axis) contributes to the re-brightening light curve at lower latitudes, driving the flux level smoothly to greater values (e.g. \citealt{2017arXiv171203237L}).

The early re-brightening portion of the light curve comes from the off-axis mildly relativistic material moving toward the observer along the line of sight and is essentially ``on-axis'' emission with respect to the observer. At $\theta_{\rm obs} = 0.4$, the slope of the re-brightening light curve from the $0.2 < \theta < 0.6$ region is moderate $\sim 1.3$ (top right panel) while the slope of the re-brightening light curve from the $0 < \theta < 0.2$ region is significantly larger $\sim 3$. The difference in the slope value results from whether the light curve is observed ``on-axis'' or ``off-axis'' with respect to the line of sight.  Observers located outside of the beaming cone of the relativistic shell $\theta_{\rm{obs}} > 1/\Gamma$, will see an ``off-axis'' light curve. The observed ``off-axis'' light curve should rise faster than $t_{\rm obs}^3$ \citep{2018arXiv180109712N}. For the GW170817 BNS merger event, the fact that the observed multi-band light curve is much shallower, scaling as $F_{\nu}\propto t_{\rm obs}^{0.78}$, implies that ``on-axis'' emission was always observed for this event \citep{2018arXiv180109712N}. 

For the structured jet model, that ``on-axis'' emission comes from the mildly relativistic sheath at an off-axis angle $\theta_{\rm{obs}}\sim 20^{\degree}$. The energy-averaged Lorentz factor at this angle is around $\Gamma \sim 3$ (Figure \ref{fig:Structure_SJet}, lower right panel), in agreement with the analytical constraint, $\Gamma\sim 1.5 - 7$, from \cite{2018arXiv180109712N}. When the central ultra-relativistic core decelerates and become ``on-axis'', the light curve stops increasing and \emph{smoothly} turns over. The peak flux is determined by the central relativistic core. This is consistent with the peak time and the peak flux estimates discussed in Section \ref{sec:lcan}. 

A similar re-brightening feature occurs in the observations of short GRBs (see e.g. \citealt{2006A&A...454..113C,2015ApJ...807..163G}), long GRBs (e.g. \citealt{2010MNRAS.402...46M}), and X-ray Flashes (e.g. \citealt{2004ApJ...605..300H}). The analysis of the off-axis light curve made here may provide an alternative interpretation for these re-brightening events. 

%What's more, we could suspect that some of the observed burst events with an early rapidly decaying afterglow %could have a rising signal, which may have been ignored.

% =====================================================================
% =====================================================================
\section{Possibility of a non-thermal X-ray ``merger flash''} \label{sec:early}

\begin{figure*}[!ht]
  \centering  
  \subfloat[\label{fig:spectra_off_axis_complete_n1e4}]{
    \includegraphics[clip,width=1.0\columnwidth]
    {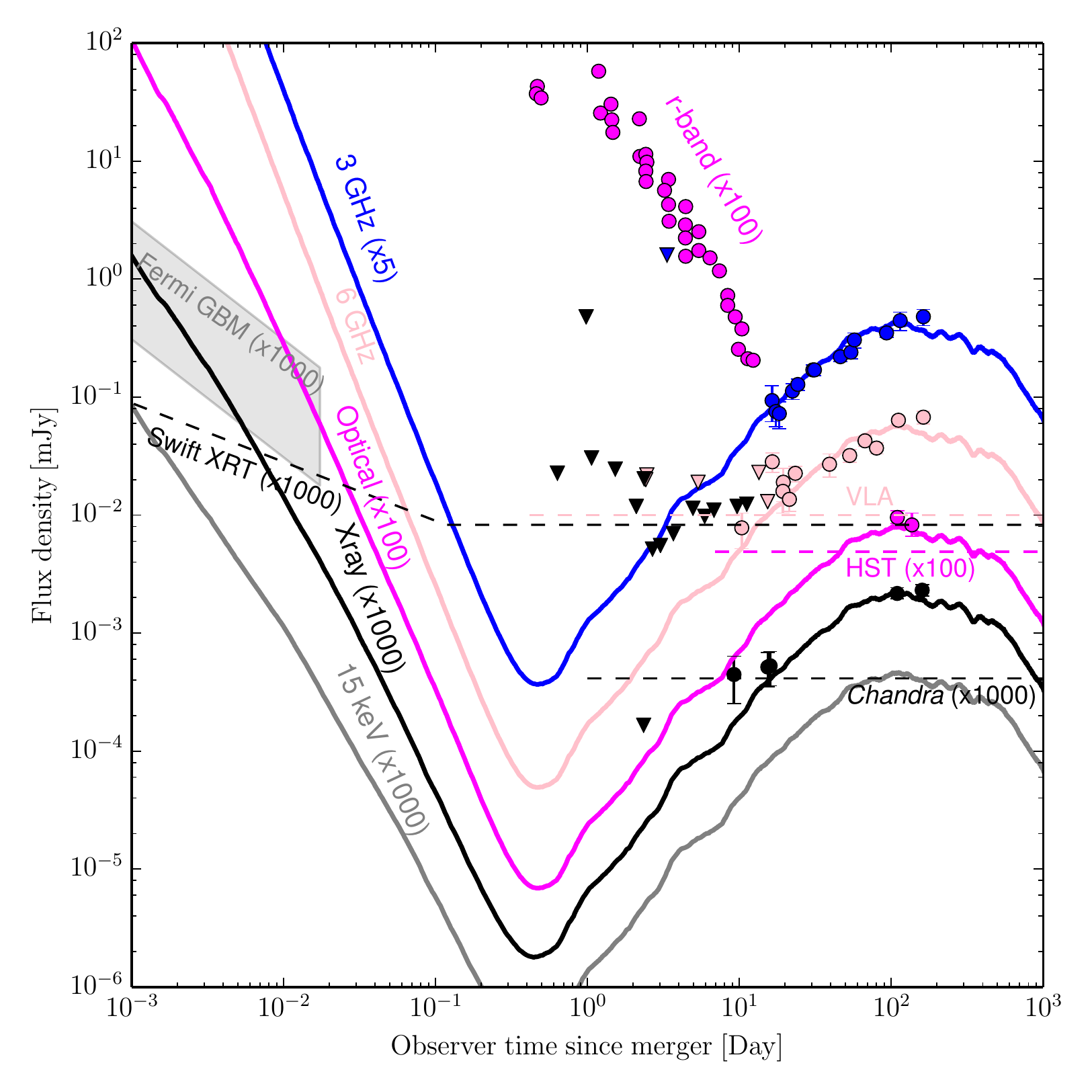}
  }
  \subfloat[\label{fig:spectra_off_axis_complete_n1e5}]{
    \includegraphics[clip,width=1.0\columnwidth]
    {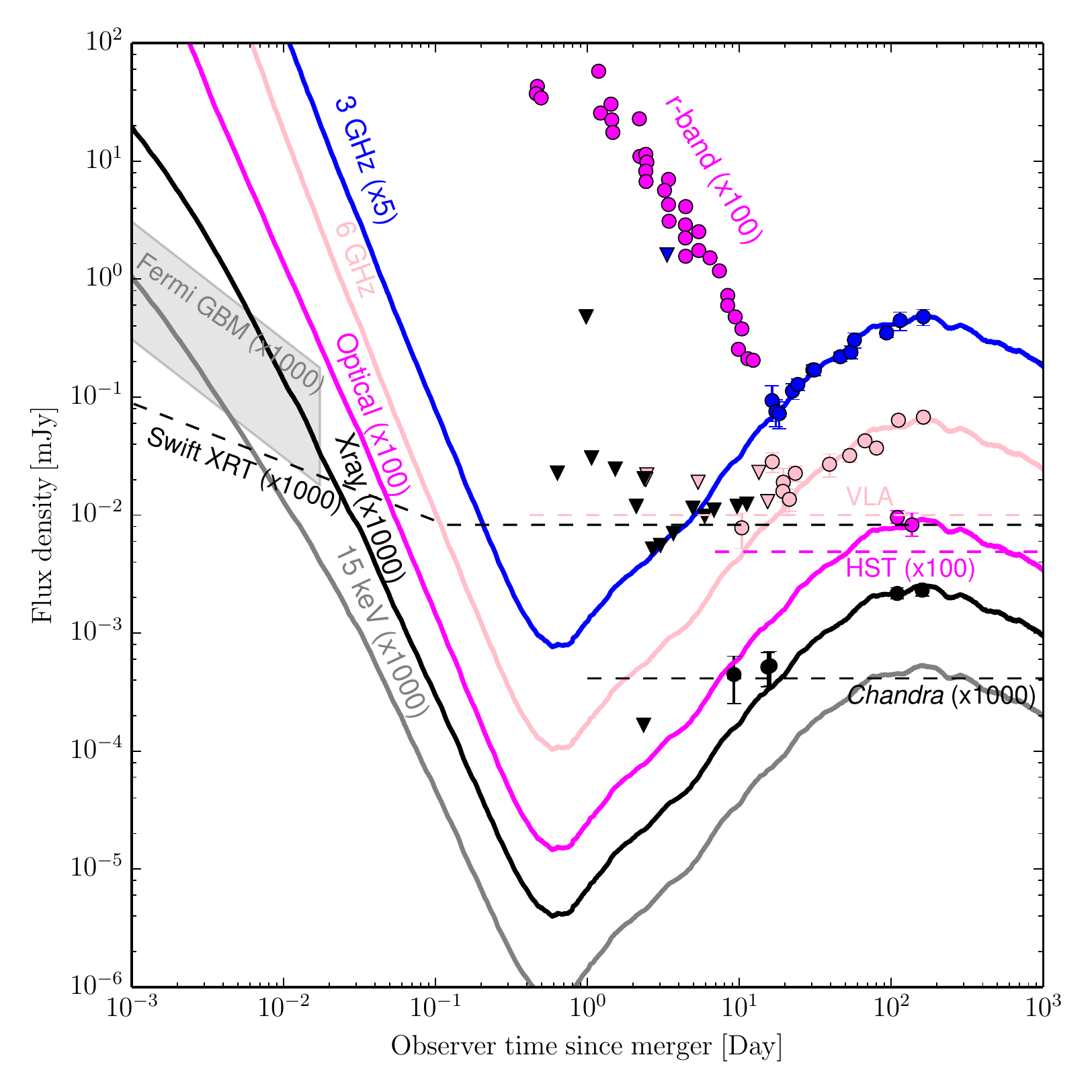}
  }
  \caption{\emph{Narrow engine model} --- Similar to Figure \ref{fig:spectra_off_axis}. 
    The plot shows two sets of fitting light curves calculated from the simulation of a structured jet propagating in an uniform ISM  environment. In the left panel (a), the ISM density is $\unit[10^{-4}]{cm^{-3}}$. The fitting radiation parameter values are $\theta_{\rm obs} = 0.34$ ($19.5^{\degree}$), $\epsilon_e = 0.02$, $\epsilon_B = 10^{-3}$, and $p = 2.16$. In the right panel (b), the ISM density is $\unit[10^{-5}]{cm^{-3}}$. The fitting  radiation parameter values are $\theta_{\rm obs} = 0.3 (17^{\degree})$, $\epsilon_e = 0.1$, $\epsilon_B = 5 \times 10^{-4}$, and $p=2.16$. The light curves shown here are not smoothed by Savitzky-Golay and are extended to the early observer time starting from $\sim \unit[1]{min}$. One added light curve (grey solid line) represents the flux of hard X-ray ($\unit[15]{keV}$), which is not (left panel) or barely (right panel) detectable by Swift BAT and Fermi GBM. The thresholds of their sensitivity are shown in the shaded region. We present the fitted multi-frequency light curves along with the detection limits of Swift-XRT (black), \emph{Chandra} (black), HST (magenta), and VLA (pink). Swift-XRT is a promising tool to detect the early declining light curve if the BNS merger site has been found within an hour or so. We calculate the flux assuming the luminosity distance is $\unit[40]{Mpc}$.}
  \label{fig:spectra_off_axis_complete}
\end{figure*}

\begin{figure*}[!ht]
  \centering  
  \subfloat[\label{fig:comparison_1keV}]{
    \includegraphics[clip,width=1.0\columnwidth]
    {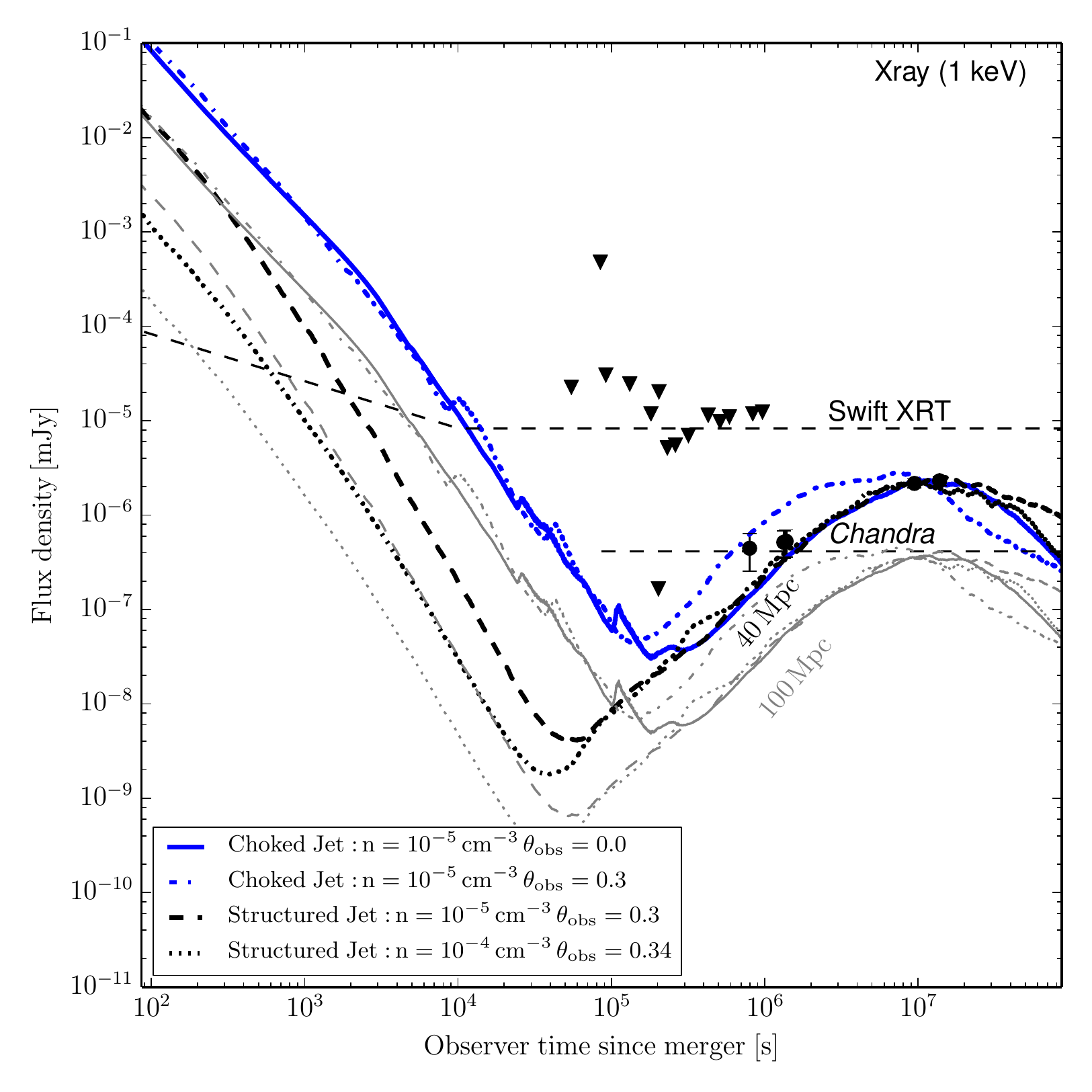}
  }
  \subfloat[\label{fig:comparison_15keV}]{
    \includegraphics[clip,width=1.0\columnwidth]
    {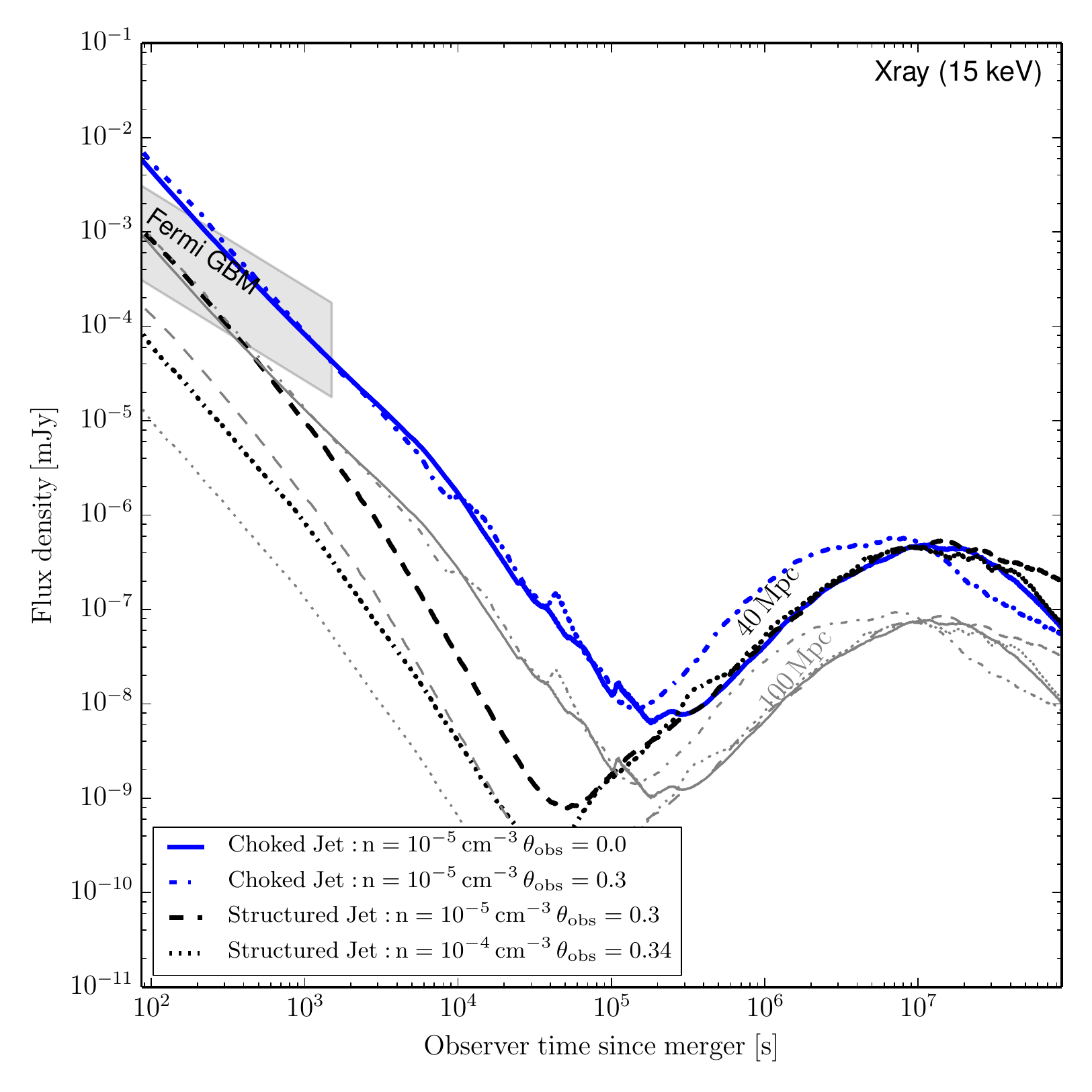}
  }
 \caption{The comparison of the soft X-ray (1keV) light curve (left panel) and hard X-ray (15 keV) light curve (right panel) calculated from the two structured jet (narrow engine) simulations (thick black lines) and one choked jet (wide engine) simulation (thick blue lines) performed in this study. The fitting radiation parameter values are listed in the captions of Figures \ref{fig:spectra_off_axis} and \ref{fig:lc_off_axis_CJet}. For the choked jet model, we also include the X-ray light curve at an off-axis viewing angle $\theta_{\rm{obs}}=0.3$ (blue dot-dashed line) with the same microphysical parameter values adopted in the calculation of on-axis light curves. Two sets of comparison light curves are shown in the plot, corresponding to the same source observed at $40\,\rm{Mpc}$ (thick lines) and $100\,\rm{Mpc}$ (thin grey lines), respectively. The flux magnitude of the early declining light curve from the on-axis choked jet model is significantly higher than off-axis structured jet models. The turning point in the light curve depends on the engine model, the viewing angle and the ISM density.}
     \label{fig:model_lc_comparison}
\end{figure*}

It has been recognized \citep{2017ApJ...834...28N,2018ApJ...855..103P}
that shock-heating of the merger cloud by the relativistic jet may
produce an observable thermal optical or UV flash at early times
(minutes to hours) following the merger. Our hydrodynamic simulations
are in overall agreement with this picture. We observe significant
heating of the merger ejecta, resulting from a strong shock wave that
is launched when the relativistic jet emerges from the
cloud. The latest shock heating
  episode occurs at high optical depth, roughly $2
\times \unit[10^{11}]{cm}$ from the merger center (see Section
\ref{sec:sjet}). %% A majority of the shock-heated merger ejecta expands
%% laterally, and is expelled significantly off-axis from the
%% relativistic core of the jet, forming the angular structure also
%% discussed in Section \ref{sec:sjet}.
The newly shock-heated material reaches
temperature on the order of $\unit[10^{7}]{K}$, and accelerates to a Lorentz
factor $\geq 2$. Here we make the thermal
equilibrium assumption, and calculate the temperature according to
$T^{\prime}=(3p^{\prime}/a)^{1/4}$, where $p^{\prime}$ is the comoving
pressure of the fluid, $a$ is the radiation constant. This material becomes optically thin after
expanding to a radius $\sim \unit[10^{12} - 10^{13}]{cm}$, at which
point the temperature has decreased adiabatically to $\sim \unit[10^4]{K}$. If radiating thermally, this material would produce a
detectable UV flash.

Here we discuss the possibility that this newly shock-heated thin layer of relativistic
  material ($\Gamma \geq 2 $) with total mass of $\sim 4\times 10^{-6}M_{\odot}$ might instead
radiate non-thermally. This would shift the emission to higher
energies, potentially rendering it detectable by Swift XRT or even
Fermi GBM, as well as future proposed wide-field X-ray
detectors. Non-thermal emission from the shock-heated merger ejecta could
be easily differentiated from the early afterglow signal, because it
is declining, whereas emission from the
external shock is brightening.

% =====================================================================
\subsection{Detectability of an X-ray merger flash} \label{sec:detectability}
The early declining light curves are computed with the synchrotron radiation model  applied to the optically thin shock-heated merger ejecta. The early emission (hereafter a \emph{merger flash}) decreases in time because of adiabatic cooling of the previously accelerated electrons. The flash is overtaken in all wave bands by rising synchrotron radiation from the external shock after roughly a day. 

For the GW170817 BNS merger event, any early declining phase has been missed. The optical flux of the early synchrotron radiation is faint compared to the observed kilonova optical data (e.g., R-band). Early X-ray emission at several hours is below the instrument detection limit of \emph{Chandra}. These are shown in Figure \ref{fig:spectra_off_axis_complete}, which displays the detection limits of various instruments along with the observational data, and two sets of fitting light curves. 

In X-ray, the late-XRT observations use a detection limit of $\unit[2 \times 10^{-14}]{erg \, cm^{-2} \, s^{-1}}$ (for a $\unit[10]{ks}$ exposure). For early-XRT, the detection limit is assumed to scale with the square root of the exposure time. For \emph{Chandra}, we adopt a constant detection limit of $\unit[10^{-15}]{erg \, cm^{-2} \, s^{-1}}$. In Figure \ref{fig:spectra_off_axis_complete}, the X-ray detection limits have been converted to the flux limits in units of $\unit[]{mJy}$ assuming the default X-ray photon energy is $\unit[1]{keV}$. In the optical, R-band imaging detection limit for HST is set to 27. In the radio, the detection limit of VLA is set to $\unit[10]{\mu Jy}$, assuming a $\unit[10]{h}$ reaction time. 

The associated early X-ray light curve would be detectable by Swift XRT until about 30 minutes following the GRB prompt emission. The hard X-ray light curve ($\unit[15]{keV}$) becomes not or barely detectable by Swift BAT and Fermi GBM after one minute \footnote{We take the $\unit[15 - 150]{keV}$ band sensitivity of Swift BAT and Fermi GBM and divide it by the corresponding frequency of photon energy $\unit[15]{keV}$ and $\unit[150]{keV}$. This gives an approximation to the flux detection limits of these two instruments.}. However, under favorable conditions, the detection of the early declining afterglow in radio, optical, and X-ray at large off-axis angles may be possible for nearby BNS mergers.

%One important ambiguity associated with GRB170817A BNS merger is the profile of the emerged outflow and how are %we supposed to distinguish which is the correct model based on current observations.
%% In sections \ref{sec:sjet} and \ref{sec:cjet}, both the off-axis ultra-relativistic structured jet and the on-axis mildly-relativistic quasi-spherical jet could produce light curves that nicely match late-time afterglow  observations.  
 %We propose that the early X-ray flash will shed light on the model
 %selection and provide insights for future detections of such
 %events.
 
% =====================================================================
\subsection{Distinguishing between successful jet and quasi-isotropic explosion from early X-ray emission} \label{sec:distinguish}
%
%% As seen in figures \ref{fig:spectra_off_axis} and \ref{fig:lc_off_axis_CJet} two dynamical models are capable of producing the late afterglow emission of GRB170817A. However, our results indicate that the sub-day emission may distinguish between these models for this event and for future BNS mergers. Figure \ref{fig:model_lc_comparison} shows the complete X-ray light curve calculated from the two narrow engine simulations with different ISM densities as well as one ``wide engine'' simulation model. The late non-thermal light curves from all three simulations match the late afterglow observations well (see Sections \ref{sec:sjet} and \ref{sec:cjet}). However, the associated sub-day light curves differ significantly from one another in their magnitude and shape.

As seen in Figures \ref{fig:spectra_off_axis} and
\ref{fig:lc_off_axis_CJet}, both the narrow and wide engine models are
capable of producing the late ($t_{\rm obs} \gtrsim 1$ day) afterglow
emission of GRB170817A. However, the
non-detection of hard X-ray ($15\,\rm{keV}$) emission following GRB170817A by Fermi GBM on
the minute timescale may disfavor the wide engine model. Because even if
seen at $\sim 20^\circ$ off axis, the quasi-isotropic explosion would
have been detected by GBM at $\sim \unit[15]{keV}$ for $\sim$ minutes
as shown in the right plot of Figure \ref{fig:model_lc_comparison}. In the wide jet
scenario, resulting in a quasi-isotropic explosion, fitting the
observed late afterglow light curve with a lower density ambient
medium requires larger values of both $\epsilon_B$ and
$\epsilon_e$. Such high values would place the X-ray merger flash
within the detection threshold of GBM, in disagreement with the
$\lesssim \unit[2]{s}$ duration of the detected short GRB
signal. Non-detection of the X-ray merger flash at one minute
  by GBM also favors the higher density ($n \sim
  \unit[10^{-4}]{cm}^{-3}$) over a lower density ($n \sim
  \unit[10^{-5}]{cm}^{-3}$) ISM environment.

GW170817 occurred in a part of the sky not accessible to Swift due to
Earth occultation. However, had this event been accessible to the
Swift satellite and XRT had slewed to its location within minutes. We
show in the left plot of Figure \ref{fig:model_lc_comparison} that XRT could have
detected a declining merger flash at $\sim \unit[1]{keV}$ lasting for
$\sim$minutes.

Future BNS merger detections are expected to occur more frequently at
larger distances, $\gtrsim \unit[100]{Mpc}$. Had GW170817 occurred
at that distance, rather than $\unit[40]{Mpc}$, the late X-ray rebrightening signal might
  not be detected by \emph{Chandra}. Therefore, it is important to
understand the rapidly fading merger flash or what other types
of electromagnetic transients that might be detectable from BNS mergers at larger distances.

\subsection{Applicability of the synchrotron emission model} \label{sec:viability}

The emission model used to create the light curves in
Figures \ref{fig:spectra_off_axis_complete} and \ref{fig:model_lc_comparison} assumes the presence of synchrotron radiating non-thermal electrons. For it to be applicable, we require a mechanism to produce and sustain the non-thermal electron population, $\epsilon_e \lesssim 0.1$. We must also invoke the presence of magnetic energy at the level $\epsilon_B \sim 10^{-2}$.

The presence of non-thermal electrons in the outflow can be supplied by shocks or magnetic reconnection. For example, the internal (sub-photospheric) shock at $\sim \unit[2 \times 10^{11}]{cm}$ might enable first-order Fermi-acceleration that would supply non-thermal electrons. Another possibility is that particles are accelerated by reconnection of residual magnetic field in the plasma outflowing from the merger sight.

The presence of magnetic energy at the level $\epsilon_B \sim
10^{-2}$, assumed in our synchrotron radiation modeling, is justified by the presence of shocks, or by magnetic dynamo activity around the central engine.
Indeed, sub-equipartition level magnetic fields are expected to be produced downstream of the internal shock via the Weibel instability (although this depends on uncertain kinetic physics of radiation mediated shocks). Magnetic energy might also exist in the neutron star merger ejecta from either the pre-merger neutron star magnetic field, or dynamo amplification during the merger itself \citep{2013ApJ...769L..29Z}. Although magnetic energy density decreases as the merger ejecta expands, $\epsilon_B$ does not evolve significantly. This is because the energy density of the tangential magnetic field ($B_\phi$ and $B_\theta$) in the coasting shocked cloud decreases like $r^{-2}$, while the gas internal energy decreases like $r^{-2 \gamma}$ where the adiabatic index is $\gamma \ge 4/3$. Therefore under expansion alone, $\epsilon_B$ either stays the same or marginally increases with radius.

%% \begin{figure*}[!ht]
%%   \centering
%%   \includegraphics[width=0.65\textwidth]{lc_observation_fit_EOS.pdf}
%%   \caption{\emph{Narrow Engine Model} --- The fitting on-axis light curve calculated from the numerical simulation of the mildly relativistic quasi-spherical outflow propagating in an ISM environment with different equation of state(EOS). The solid lines represent the original simulation with the ideal gas EOS. The dash-dotted lines are calculated from identical simulation with RC EOS. \label{fig:lc_off_axis_EOS}
%%     }
%% \end{figure*}

\section{Conclusion and Discussion} \label{sec:conclusion}
In this study, we have presented relativistic hydrodynamic
simulations to explore the dynamics and radiative signatures of
merging neutron star outflows. We have focused our modeling on two
primary scenarios, dubbed the narrow and wide engine
models. The narrow jet engine penetrates the debris cloud surrounding the merger site, and propagates successfully into the circum-merger
medium. This successful jet may drive a classical short
gamma-ray burst if viewed on-axis. In contrast, the
wide jet engine fails to break out of the merger cloud, and instead drives a quasi-spherical shock through the cloud and into the surrounding medium.

Both the narrow and wide engine models can explain the afterglow of
GRB170817A, including observations through $\sim \unit[200]{days}$
after the GW signal (see Figures \ref{fig:spectra_off_axis} and \ref{fig:lc_off_axis_CJet}). We find that in both scenarios, the jet develops an angular structure as a result of its interaction with the merger ejecta cloud. Both models predict the afterglow light curve to begin decaying after $\sim 200$ days, in a similar manner. Thus, upcoming observations of the late afterglow emission may not resolve the question of which scenario was the case for the GW170817 BNS merger event. Similar conclusions are made in \citep{2018ApJ...856L..18M, 2018arXiv180109712N}.

However, as we discussed in Section \ref{sec:distinguish}, we surmise
that non-detection of longer-lived ($\sim \unit[]{minutes}$) hard
X-ray emission by GBM disfavors the wide engine model. Instead, the
narrow engine model is favored because it can produce well-fitted late
afterglow light curves without over-predicting the magnitude of the
early X-ray flash. As discussed in Section \ref{sec:viability}, these
conclusions are dependent on the presence of synchrotron radiating
non-thermal electrons in the mildly relativistic shock-heated cocoon. Hence, the detection of an X-ray merger flash is potentially
valuable as a probe of previously unexplored plasma conditions. In
particular, its existence would indicate that either electrons are
accelerated by sub-photospheric, radiation-mediated shocks, or by
sustained dissipation of magnetic energy as the shell expands.

Previous studies have also considered the radiative signatures of
structured jets \citep{2017ApJ...848L...6L, 2017arXiv171203237L,
  2018MNRAS.473L.121K, 2018arXiv180102669L, 2018MNRAS.tmpL..60T,
  2018MNRAS.tmp.1056L}. In this work we have conducted simulations
starting from the scale of the engine and continuing self-consistently
to the afterglow stage. These engine-to-afterglow simulations reveal
that jet structures that are consistent with the observations are a
natural consequence of the hydrodynamical interaction of the jet with
the ejecta cloud of merging binary neutron stars.

\acknowledgements
We thank Andrei Gruzinov, Brian Metzger, Geoffrey Ryan and
  Yiyang Wu for helpful comments and discussions. This research was supported in part by the National Science Foundation under Grant No. AST-1715356.

\clearpage

\bibliography{sgrb}

\end{document}